\documentclass[fleqn,usenatbib]{mnras}

\usepackage{newtxtext,newtxmath}

\usepackage[T1]{fontenc}
\usepackage{ae,aecompl}

\usepackage{graphicx}	
\usepackage{amsmath}	
\usepackage{hyperref}

\def\apj{ApJ}
\def\pasp{PASP}

\def\araa{ARA\&A}
\def\aap{A\&A}
\def\apjl{ApJL}
\def\apjs{ApJS}
\def\mnras{MNRAS}
\def\aj{AJ}
\def\nat{Nature}

\def\icarus{Icarus}

\hypersetup{
  draft=false,
	pdftoolbar=false,
	pdfmenubar=true,
	pdfstartview={FitP},
	breaklinks=false,
	frenchlinks=false,
  colorlinks=true,
  linkcolor=red,
  citecolor=blue,
  filecolor=cyan,
  urlcolor=magenta,
  pdftitle={},
	pdfsubject={},
	pdfauthor={Elspeth K. H. Lee}
  }

\title[Exoplanet Gas Giants with \textsc{Exo-FMS}]{Simulating gas giant exoplanet atmospheres with \textsc{Exo-FMS}: \\
Comparing semi-grey, picket fence and correlated-k radiative-transfer schemes.}

\author[Lee et al.]{
Elspeth K. H. Lee$^{1}$,
Vivien Parmentier$^{2}$,
Mark Hammond$^{3}$,
Simon L. Grimm$^{1}$,
\newauthor
Daniel Kitzmann$^{1}$,
Xianyu Tan$^{2}$,
Shang-Min Tsai$^{2}$ \newauthor
and Raymond T. Pierrehumbert$^{2}$
\\
$^{1}$Center for Space and Habitability, University of Bern, Gesellschaftsstrasse 6, CH-3012 Bern, Switzerland \\
$^{2}$Atmospheric, Oceanic \& Planetary Physics, Department of Physics, University of Oxford, Oxford OX1 3PU, UK \\
$^{3}$Department of the Geophysical Sciences, The University of Chicago, Chicago, IL 60637, USA
}

\date{Accepted 22 Jun 2021. Received YYY; in original form ZZZ}

\pubyear{2021}

\begin{document}
\label{firstpage}
\pagerange{\pageref{firstpage}--\pageref{lastpage}}
\maketitle

\begin{abstract}
Radiative-transfer (RT) is a fundamental part of modelling exoplanet atmospheres with general circulation models (GCMs).
An accurate RT scheme is required for estimates of the atmospheric energy transport and for gaining physical insight from model spectra.
We implement three RT schemes for \textsc{Exo-FMS}: semi-grey, non-grey `picket fence', and real gas with correlated-k.
We benchmark the \textsc{Exo-FMS} GCM using these RT schemes to hot Jupiter simulation results from the literature.
We perform a HD 209458b-like simulation with the three schemes and compare their results.
These simulations are then post-processed to compare their observable differences.
The semi-grey scheme results show qualitative agreement with previous studies in line with variations seen between GCM models.
The real gas model reproduces well the temperature and dynamical structures from other studies.
After post-processing our non-grey picket fence scheme compares very favourably with the real gas model, producing similar transmission spectra, emission spectra and phase curve behaviours.
\textsc{Exo-FMS} is able to reliably reproduce the essential features of contemporary GCM models in the hot gas giant regime.
Our results suggest the picket fence approach offers a simple way to improve upon RT realism beyond semi-grey schemes.
\end{abstract}




\section{Introduction}
\label{sec:intro}

The three dimensional (3D) atmospheric properties of hot Jupiters give rise to each of their unique spectral properties observed by contemporary telescopes and instrumentation technology.
Understanding the 3D structure of these atmospheres and modelling them accurately is vital for a physical interpretation of their atmospheric dynamical and chemical characteristics.
Recent studies have also shown the importance of considering the multi-dimensional nature of atmospheres when interpreting observational data \citep[e.g.][]{Feng2016, Dobbs-Dixon2017, Blecic2017, Caldas2019, Irwin2020, Taylor2020}.
These studies suggest that neglecting the 3D nature of the atmosphere can lead to biased or inaccurate results when interpreting the thermal and molecular properties of the atmosphere.

During the operational lifetime of JWST it is expected that several phase curves observations of exoplanet atmospheres will be performed, providing wide infrared wavelength, detailed global atmospheric information.
The Transiting Exoplanet Community Early Release Science (ERS) Program for JWST \citep{Bean2018, Venot2020} plans to observe phase curves of the hot Jupiter WASP-43b.
In addition, the Transiting Exoplanet Survey Satellite (TESS) \citep{Ricker2014}, CHaracterising ExOPlanet Satellite (CHEOPS) \citep{Broeg2013} and the high altitude ballon mission EXoplanet Climate Infrared TElescope (EXCITE) \citep{Nagler2019} are able to observe several phase curves of hot Jupiters in the optical and infrared wavelength regimes.
Recently, TESS data on the phase curves of WASP-18b \citep{Shporer2019} and WASP-19b \citep{Wong2020b}, WASP-121b \citep{Daylan2021, Bourrier2020}, WASP-100b \citep{Jansen2020} and KELT-9b \citep{Wong2020} have been published.
As more phase curve data of nearby characterisable planets becomes available, a larger census of the dynamical properties of hot Jupiter atmospheres, as well as their cloud coverage through dayside albedo constraints \citep[e.g.][]{Cahoy2010, Barstow2014,Munoz2015,Parmentier2016,Batalha2019,Mayorga2019} will become possible.
The Atmospheric Remote-sensing Infrared Exoplanet Large-survey (ARIEL) \citep{Tinetti2016} telescope is also scheduled to perform numerous phase curve observations of hot gas giant planets \citep{Charnay2021}.

High resolution spectroscopy observations have also helped reveal the rich chemical inventory of hot Jupiter atmospheres \citep[e.g.][]{Seidel2019, Ehrenreich2020, Merritt2020, Pino2020}.
3D modelling efforts have shown that considering the 3D structures in detail affects the interpretation of high resolution transmission spectra results \citep[e.g.][]{Kempton2012,Flowers2019}.
Possible improvements in the molecular detection significance for high resolution emission spectra using GCM model output as a template has also been reported in \citet{Beltz2021}.

The 3D modelling of hot Jupiter (HJ) atmospheres using large scale simulation platforms has become an important toolkit in understanding the physical processes that give rise to the observational characteristics of these planets \citep{Showman2020}.
Several groups have examined hot Jupiter (HJ) and warm Neptune atmospheres with 3D General Circulation Models (GCMs) or Radiative-Hydrodynamic (RHD) models, for example, \citet{Showman2009, Rauscher2010, Heng2011, Dobbs-Dixon2013, Mayne2014, Charnay2015a, Mendonca2016}.
Each of these models generally produce qualitatively similar dynamical structures \citep{Heng2015}, despite many differences in their implementations, the level of simplification from the Navier-Stokes equations, numerics, 3D grid structure, vertical coordinate system to radiative-transfer scheme.

Radiative-transfer (RT) is a key component in GCM models, controlling the heating and cooling in the atmosphere which majorly impacts the dynamical properties of the atmosphere.
Due to their importance, significant effort has been made developing a broad spectrum of RT schemes in the literature.
Equilibrium relaxation or simplified RT approaches such as Newtonian relaxation \citep[e.g.][]{Showman2008, Rauscher2010, Heng2011, Mayne2014, Carone2020} and multi-band grey or non-grey schemes of various flavours \citep[e.g.][]{Heng2011b, Rauscher2012, Dobbs-Dixon2013, Mendonca2018b} are commonly used in HJ GCM modelling.
These efficient schemes have enabled easier model inter-comparison \citep{Heng2015} and larger parameter explorations \citep[e.g.][]{Komacek2016, Komacek2017, Tan2019, Tan2020} in an effort to understand dynamical regime changes.
Detailed real gas, correlated-k RT models have also been coupled to GCMs \citep[e.g.][]{Showman2009, Charnay2015a, Amundsen2016}, providing a more realistic but more costly RT solution.
Several targeted object simulations \citep{Kataria2014,Kataria2016, Amundsen2016} and parameter explorations \citep{Parmentier2016} have been performed using correlated-k modelling.
Recently, \citet{Ding2019} implemented a line-by-line RT approach in their GCM to model the atmosphere of GJ 1132b.
Similar RT schemes are also a key component for GCM and general atmospheric modelling for Solar System planets.
Many RT models have been developed, such as for Venus \citep[e.g.][]{Lebonnois2010,Mendonca2015,Mendonca2016b,Limaye2018} and for Jupiter \citep[e.g.][]{Schneider2009, Kuroda2014, Li2018, Young2019}.

In this study, we couple three different RT schemes to the 3D \textsc{Exo-FMS} GCM model, and benchmark them to several hot Jupiter models performed in the literature.
\textsc{Exo-FMS} has been previously used to model condensable rich atmospheres on terrestrial planets \citep{Pierrehumbert2016}, explore atmospheric compositions of 55 Cnc e \citep{Hammond2017} and the White dwarf - Brown dwarf short period binary WD0137-349B \citep{Lee2020}, showcasing \textsc{Exo-FMS} flexibility and applicability to model and explore a wide planetary parameter space.
We benchmark to commonly used set-ups for hot Jupiter modelling efforts and examine our semi-grey, non-grey picket fence and real gas RT models for a HD 209458b test case.
In Sect. \ref{sec:Exo-FMS} we summarise the \textsc{Exo-FMS} hydrodynamic and radiative-transfer methods.
In Sect. \ref{sec:semi-grey} we present benchmarking results using the semi-grey RT scheme.
In Sect. \ref{sec:semi-grey2}, \ref{sec:non-grey} and \ref{sec:spectral} we present HD 209458b-like simulations using the semi-grey, non-grey picket fence and real gas RT schemes respectively.
The differences in each model are examined in Sect. \ref{sec:comp}.
Section \ref{sec:pp} presents transmission, emission and phase curve post-processing of the HD 209458b-like results and comparisons between the three models and available observational data.
Section \ref{sec:discussion} contains the discussion of our results and Sect. \ref{sec:conclusions} contains the summary and conclusions.

\section{Exo-FMS}
\label{sec:Exo-FMS}

\textsc{Exo-FMS} is a GCM adapted from the Princeton Geophysical Fluid Dynamics Laboratory (GFDL) Flexible Modelling System (FMS).
Other groups have also used the FMS framework for exoplanet studies, such as the ISCA model \citep[e.g.][]{Thomson2019} which uses a spectral dynamical core, and the \citet{Heng2011}, \citet{Heng2011b} and \citet{Oreshenko2016} studies.

In the current \textsc{Exo-FMS} version, we use the finite volume dynamical core \citep{Lin2004}.
A key advantage of this version of \textsc{Exo-FMS} is the scalability afforded by the cube sphere implementation, increasing the efficiency of computationally demanding physics such as real gas radiative-transfer, chemical kinetics and cloud formation.
The cube-sphere grid also avoids singularities at the pole regions, a common problem with lat-lon based grids.
In this study we use a resolution of C48, $\approx$ 192 $\times$ 96 in longitude $\times$ latitude.
C48 is typically higher than the C32 resolution commonly used by SPARC/MITgcm \citep[e.g.][]{Parmentier2021}, C32 was found to capture most of large scale dynamical flows of hot Jupiter atmospheres \citep[e.g.][]{Showman2009}.

Exo-FMS uses a hybrid-sigma vertical pressure grid system, where deep regions are typically spaced in log-pressure starting from a `surface' reference pressure (here 220 bar), with a gradual transition to more linear spacing occurring at lower pressure.
We generally use a 53 level grid with a lowest pressure of $10^{-6}$ bar, with more linear spacing starting at $\lesssim$ 10$^{-3}$ bar.

\subsection{Hydrodynamics}

\textsc{Exo-FMS} evolves the standard primitive equations of meteorology, extensively detailed in many textbooks and sources, for example, \citet{Holton2013}, \citet{Mayne2014}, \citet{Heng2017b} and \citet{Mayne2019}.
We note that there are indications that the use of primitive equations for the slow rotating warm Neptune regime can lead to significant errors in the wind field, especially near the equatorial regions \citep{Mayne2019}.
These appear to be due to geometric terms conventionally neglected in the primitive equations, rather than the hydrostatic approximation itself.
However, the primitive equation set assumptions are well met in the hot Jupiter regime \citep{Mayne2014}.

\subsection{Dry convective adjustment}

Dry convective adjustment is included using a layerwise enthalpy conserving scheme. Convective adjustment is given by examining layer $i$ for the criteria
\begin{equation}
T_{i} < T_{i+1}\left(\frac{p_{i}}{p_{i+1}}\right)^{\kappa},
\end{equation}
where $\kappa$ = $R_{\rm atm}$/$c_{\rm p}$, the specific gas constant divided by the specific heat capacity at constant pressure of the atmosphere.
The layer pair are then adjusted to the adiabatic gradient while conserving dry enthalpy.
This processes is iterated until no pair of layers are convectively unstable and the whole column is stable.

\subsection{Semi-grey radiative transfer}
\label{sec:grey_RT}

In a semi-grey RT scheme, the irradiation by the host star (shortwave radiation) is represented by a single `visual' (V) band, with a characteristic constant opacity.
Similarly for the internal atmospheric thermal `infrared' (IR) fluxes (longwave radiation), a single band with a characteristic opacity is also used.
It is important to note that the V and IR bands do not necessarily conform to strict visual and infrared wavelength regimes.

In a two-stream context this simplifies the RT scheme to performing two sets of calculations: the downward (and sometimes upward) flux resulting from stellar irradiation, and the downward and upward thermal band fluxes.
Below we briefly summarise the main components.

\subsubsection{Shortwave radiation}
The flux at the sub-stellar point received by the planet by its host star, $F_{0}$ [W m$^{-2}$], is given by the irradiation temperature, $T_{\rm irr}$ [K],  \citep[e.g.][]{Guillot2010}
\begin{equation}
F_{0} = \sigma T_{\rm irr}^{4} = \sqrt{\frac{R_{\star}}{a}}\sigma T_{\star}^{4},
\end{equation}
where $\sigma$ [W m$^{-2}$ K$^{-4}$] is the Stefan-Boltzmann constant, $R_{\star}$ [m] the radius of the host star, $a$ [m] the orbital semi-major axis and $T_{\star}$ [K] the effective temperature of the host star.
For an atmosphere that is purely absorbing in the shortwave band, the reference shortwave optical depth $\tau_{S_{0}}$ is calculated, assuming hydrostatic equilibrium, given by
\begin{equation}
\tau_{S_{0}} = \frac{\kappa_{v}p_{0}}{g},
\end{equation}
where $\kappa_{v}$ [m$^{2}$ g$^{-1}$] is the grey visual band opacity, $p_{0}$ [pa] the reference surface pressure and g [m s$^{-2}$] the surface gravity.
The cumulative optical depth at each vertical cell level for a constant with height opacity source is then
\begin{equation}
\tau_{i} = \tau_{S_{0}} \frac{p_{i}}{p_{0}},
\end{equation}
where p$_{i}$ [pa] is the pressure at the cell interface.
The downward shortwave flux, $F_{\downarrow S}$ [W m$^{-2}$], at each vertical level is then
\begin{equation}
F_{\downarrow S} = (1 - A_{B})\mu_{\star}F_{0}\exp(-\tau_{i}/\mu_{\star}),
\end{equation}
where $\mu_{\star}$ = $\cos(\phi)\cos(\theta)$ is the cosine angle from the sub-stellar point and A$_{B}$ a parametrised Bond albedo.

The optical wavelength scattering properties of the atmosphere are implicitly included in the A$_{B}$ parametrisation.
For a purely absorbing atmosphere, the Bond albedo would be zero, but here, in line with common practice, we introduce an imposed albedo as an approximation to the effects of scattering.
It would be straightforward and still computationally efficient to introduce a two-stream formulation including scattering for the shortwave band, but since our object here is benchmarking against published results in the simplest possible way, we have retained the commonly used expedient of forcing the atmosphere with pure direct beam absorption and top-of-atmosphere flux adjusted according to an assumed albedo.
The effects of multiple scattering would alter the vertical distribution of heating, and in particular scatter direct beam into diffuse radiation, reducing the dependence on zenith angle \citep{Pierrehumbert2010}.

\subsubsection{Longwave radiation}
\label{sec:lw}

For the longwave fluxes, we follow a simplified version of the scheme presented in \citet{Toon1989}.
Ignoring the coupling coefficients for scattering (i.e. in the w$_{0}$ $\rightarrow$ 0 limit) the two-stream formulation reduces to
\begin{equation}
\label{eq:Iup}
I_{\uparrow 1} = I_{\uparrow 2}T_{0} + 2\pi B_{2}(1 - T_{0}) + 2\pi B'[\mu - (\Delta + \mu)T_{0}],
\end{equation}
for the upward intensity, $I_{\uparrow}$ [W m$^{-2}$ sr$^{-1}$],  and
\begin{equation}
\label{eq:Idown}
I_{\downarrow 2} = I_{\downarrow 1}T_{0} + 2\pi B_{1}(1 - T_{0}) + 2\pi B'[\mu T_{0} +  \Delta + \mu],
\end{equation}
for the downward intensity, $I_{\downarrow}$ [W m$^{-2}$ sr$^{-1}$],  with $\mu$ the emission angle.
We follow the notation of \citet{Heng2018} with level 2 located below 1 in altitude, with  $\Delta$ = $\tau_{2}$ - $\tau_{1}$ $>$ 0,
\begin{equation}
T_{0} = \exp\left(-\Delta/\mu\right),
\end{equation}
the transmission function, and B$'$ = (B$_{2}$ - B$_{1}$)/$\Delta$ the linear in $\tau$, non-isothermal layer blackbody term.
For very small optical depths the above scheme can become numerically unstable, we therefore switch to an isothermal approximation when the optical depth becomes small.
When layers have $\Delta$ $<$ 10$^{-6}$ the upward and downward intensity are given by
\begin{equation}
\label{eq:Iup_iso}
I_{\uparrow 1} = I_{\uparrow 2}T_{0} + \pi (B_{2} + B_{1}) (1 - T_{0}),
\end{equation}
and
\begin{equation}
\label{eq:Idown_iso}
I_{\downarrow 2} = I_{\downarrow 1}T_{0} + \pi (B_{2} + B_{1}) (1 - T_{0}),
\end{equation}
respectivly.
For grey schemes, the blackbody intensity in the above equations is given by the wavelength integrated blackbody intensity, B = $\sigma$T$^{4}$/$\pi$ [W m$^{-2}$ sr$^{-1}$].
The flux at each level is then calculated by quadrature integration of different $\mu$ values.
In this study we use two quadrature angles, which was found to give similar quality results to higher numbers of quadrature points, Appendix \ref{app:RT_validation}, \citep[e.g.][]{Lacis1991}.

The upward and downward longwave fluxes $F_{L}$ [W m$^{-2}$] at each level are then given by
\begin{equation}
F_{\uparrow L} = \sum_{g}^{2}w_{g}\mu_{g}I_{\uparrow, g} \\
F_{\downarrow L} = \sum_{g}^{2}w_{g}\mu_{g}I_{\downarrow, g},
\end{equation}
where w$_{g}$ is quadrature weight and $\mu_{g}$ the emission angle.

The net flux into and out of the cell level is then
\begin{equation}
F_{\rm net} = F_{\uparrow L} - F_{\downarrow L} - F_{\downarrow S},
\end{equation}
with the flux from the deep interior is modelled as a net flux at the deepest level given by the internal temperature, T$_{\rm int}$ [K],
\begin{equation}
F_{\rm net, int} = \sigma T_{\rm int}^{4}.
\end{equation}

The temperature tendency due to radiative heating/cooling, $\partial$T/$\partial$t [K s$^{-1}$], is then given by
\begin{equation}
\frac{\partial T}{\partial t} = \frac{g}{c_{\rm p, air}}\frac{\partial F_{\rm net}}{\partial p},
\end{equation}
where g [m s$^{-2}$] is the surface gravity and c$_{\rm p, air}$ [J kg$^{-1}$ K$^{-1}$] the heat capacity of the air.

In the semi-grey scheme, the longwave optical depth at each cell interface is given following \citet{Heng2011}
\begin{equation}
\tau_{i} = f_{l}\tau_{L_{0}} \frac{p_{i}}{p_{0}} + (1 - f_{l})\tau_{L_{0}} \left(\frac{p_{i}}{p_{0}}\right)^{n_{L}},
\end{equation}
where $f_{l}$ is the fraction between the constant (linear term) which represents molecular and atomic line opacity and exponent term, $\tau_{L_{0}}$ the reference longwave optical depth and $n_{L}$ the pressure dependence exponent, representing collision-induced continuum opacity.

\subsection{Non-grey picket fence scheme}
\label{sec:non-grey_RT}

As an intermediate model between the semi-grey and real gas RT models we develop a non-grey `picket fence' RT approach \citep{Chandrasekhar1935} based on \citet{Parmentier2014} and \citet{Parmentier2015}.
The picket fence scheme attempts to more accurately model the propagation of radiation through the atmosphere by the use of two opacities, one representing the molecular and atomic line opacity and one representing the general continuum opacity.
The value of these opacities are linked to the local temperature and pressure dependent Rosseland mean opacity through fitting functions derived analytically in \citet{Parmentier2014,Parmentier2015} and tuned to best fit the results of a correlated-k model for each value of T$_{\rm eff}$.
In addition to this, \citet{Parmentier2015} use additional V band opacity relations to better capture the deposition of incident stellar flux.
The key concepts from \citet{Parmentier2014, Parmentier2015} applied here are the use of three V bands and two picket fence IR bands (again these bands do not strictly correspond to optical and infrared wavelength regimes).

The optical depth contribution in each layer, $\Delta\tau_{i}$, assuming hydrostatic equilibrium, is given by
\begin{equation}
\Delta\tau_{i} = \frac{\kappa_{i} (p_{i},T_{i})}{g}\Delta p_{i},
\end{equation}
where $\kappa_{i}$ [m$^{2}$ g$^{-1}$] is the opacity of the layer, now dependent on the local pressure and temperature, g [m s$^{-2}$] the gravitational acceleration and $\Delta p_{i}$ [pa] the difference in pressure between the cells lower and upper level respectively.
The value of $\kappa_{i}$ is calculated in each layer for each V and IR band as function of the local Rosseland mean opacity, $\kappa_{R}$ [m$^{2}$ g$^{-1}$], given by
\begin{equation}
\frac{1}{\kappa_{R}} = \frac{\int^{\infty}_{0}\frac{1}{\kappa_{\lambda}}\frac{dB_{\lambda}}{dT}d\lambda}{\int^{\infty}_{0}\frac{dB_{\lambda}}{dT}d\lambda},
\end{equation}
where $\kappa_{\lambda}$ [m$^{2}$ g$^{-1}$] is the wavelength dependent opacity and dB$_{\lambda}$/dT the temperature derivative of the Planck function, and opacity ratio coefficients, $\gamma_{i}$,  \citep{Parmentier2014, Parmentier2015} through the relation
\begin{equation}
\gamma_{i} = \frac{\kappa_{i}}{\kappa_{R}}.
\end{equation}

The downward V band flux is now calculated as a sum of the downward stellar flux calculations with opacities $\kappa_{V,i}$
\begin{equation}
F_{\downarrow S} = (1 - A_{B})F_{0}\mu_{\star} \sum_{b}^{N_{b}}\beta_{V, b}\exp(-\tau_{i, b}/\mu_{\star}),
\end{equation}
where $N_{b}$ is the number of V bands (here three), $\beta_{V, b}$ the fraction of stellar flux in band $b$ (here 1/3) and A$_{B}$ the Bond albedo.
The value of A$_{B}$ can be calculated from the fitting function presented in \citet{Parmentier2015} or given as a parameter.

For the longwave flux calculations, the picket fence model considers two bands, with opacities $\kappa_{1}$ and $\kappa_{2}$, each with a fraction $\beta$ of the integrated blackbody flux.
Equations \ref{eq:Iup} and \ref{eq:Idown} are performed twice, one with opacity $\kappa_{1}$ and $\beta$B and one with $\kappa_{2}$ and (1 - $\beta$)B, the flux of both bands are then summed to find the net flux.

Following \citet{Parmentier2014,Parmentier2015} we define the effective temperature, T$_{\rm eff}$ [K], of an individual vertical profile as
\begin{equation}
T_{\rm eff}^{4} = T_{\rm int}^{4} + (1 - A_{B})\mu_{\star}T_{\rm irr}^{4},
\label{eq:Teff}
\end{equation}
where $\mu_{\star}$ is the cosine angle of the column from the sub-stellar point, A$_{B}$ the Bond albedo, T$_{\rm irr}$ [K] the substellar point irradiation temperature and T$_{\rm int}$ [K] the internal temperature.
For nightside profiles T$_{\rm eff}$ = T$_{\rm int}$.

On a practical level, the scheme operates as follows:
\begin{enumerate}
\item Calculate the Bond albedo following the \citet{Parmentier2015} expression, with T$_{\rm eff}$ assuming $\mu_{\star}$ = 1/$\sqrt{3}$.
\item Calculate the $\gamma_{v1}$, $\gamma_{v2}$, $\gamma_{v3}$, $\beta$, $\gamma_{p}$ and R = $\kappa_{1}/\kappa_{2}$ constants from the profile  T$_{\rm eff}$ for each column's $\mu_{\star}$, Eq. \ref{eq:Teff}, following the fitting coefficient tables in \citet{Parmentier2015} and definitions in \citet{Parmentier2014}.
\item Find the IR band Rosseland mean opacity, $\kappa_{R}$ (p,T), in each layer from the \citet{Freedman2014} fitting function.
\item Find the three V band opacities in each layer using the $\gamma_{v1}$, $\gamma_{v2}$, $\gamma_{v3}$ and $\kappa_{R}$ relationships.
\item Find the two IR band opacities in each layer using the $\kappa_{1}, \kappa_{2}$ and $\gamma_{P}$, $\beta$, R relationships.
\item Calculate the vertical optical depth structure and perform the two-stream calculations for each V and IR band.
\end{enumerate}
Each column in the GCM has a constant T$_{\rm eff}$ as defined by this scheme, so the \citet{Parmentier2015} parameters need only be calculated once at the start of the simulation.
The $\gamma_{v1}$, $\gamma_{v2}$, $\gamma_{v3}$, $\beta$, $\gamma_{p}$ and R = $\kappa_{1}/\kappa_{2}$ constants were parametrised in \citet{Parmentier2015} to match the T-p profiles of the \citet{Fortney2005} and collaborators 1D radiative-convective equilibrium (RCE) model across a wide variety of T$_{\rm eff}$.
In essence, the these parameters produce opacity structures, combined with the dry convective adjustment scheme, that tend each column to the analytical RCE T-p profiles given by \citet{Parmentier2015}, with the energy transport timescales from the atmospheric dynamics giving rise to the non-equilibrium behaviour of the 3D temperature structures.

Overall, this picket fence scheme attempts to add more realism than semi-grey case, at the cost of now performing 3 shortwave flux calculations and 2 longwave flux calculations as well as calculating Rosseland mean opacities.
This model is therefore typically $\approx$2-3 times more computationally expensive than the semi-grey calculation, but presents a more physical way of representing the thermal structure of the atmosphere as opacities are now able to adapt to the local temperature and pressure variations.
In addition, collision-induced absorption opacities are included in the Rosseland mean calculation resulting in a more accurate representation of the pressure dependence of the opacity structure without relying on tuning the pressure dependent exponent in the semi-grey scheme.

\subsection{Real gas radiative transfer}
\label{sec:spec_RT}

\begin{table}
\centering
\caption{Opacity sources and references used in the GCM correlated-k scheme and \textsc{cmcrt} post-processing. `NU' denotes `not used'.}
\begin{tabular}{c c}  \hline \hline
Opacity Source & GCM model / CMCRT post-processing   \\ \hline
Line &  Reference \\ \hline
Na  &  \citet{Kurucz1995}/\citet{NISTWebsite}   \\
K   &  \citet{Kurucz1995}/\citet{NISTWebsite}  \\
H$_{2}$O   &  \citet{Polyansky2018}   \\
CH$_{4}$   &  \citet{Yurchenko2017}  \\
C$_{2}$H$_{2}$   &  \citet{Chubb2020}   \\
NH$_{3}$   &  \citet{Coles2019}   \\
CO   &  \citet{Li2015}   \\
CO$_{2}$    &  \citet{Rothman2010}   \\
H$_{2}$S   & \citet{Azzam2016}/NU \\
HCl   & \citet{Gordon2017}/NU \\
HCN   & \citet{Harris2006, Barber2014}/NU \\
SiO & \citet{Barton2013}/NU \\
PH$_{3}$ & \citet{Sousa_Silva2015} \\
TiO & \citet{McKemmish2019}/NU \\
VO & \citet{McKemmish2016}/NU \\
MgH & \citet{GharibNezhad2013}/NU \\
CaH & \citet{Bernath2020}/NU \\
TiH & \citet{Bernath2020}/NU \\
CrH & \citet{Bernath2020}/NU \\
FeH & \citet{Bernath2020}/NU \\ \hline
Continuum  & \\ \hline
H$_{2}$-H$_{2}$  &  \citet{Karman2019} \\
H$_{2}$-He   & \citet{Karman2019} \\
H$_{2}$-H   & \citet{Karman2019} \\
H-He   & \citet{Karman2019} \\
H$^{-}$ & \citet{John1988} \\
He$^{-}$  & \citet{Kurucz1970} \\ \hline
Rayleigh scattering & \\ \hline
H$_{2}$ & \citet{Dalgarno1962} \\
He  & \citet{Thalman2014} \\
H  & \citet{Kurucz1970} \\
e$^{-}$ & Thompson scattering\\ \hline
\end{tabular}
\label{tab:line-lists}
\end{table}

For our real gas radiative scheme, we use the same two-stream formulations as in Sect. \ref{sec:lw} for a fair comparison to the grey and picket fence schemes, with modifications to use the 30 wavelength band structure from \citet{Showman2009}.
We produce k-coefficient tables using the high resolution opacities from HELIOS-K \citep{Grimm2015,Grimm2021} with Na and K opacities from \citet{Kitzmann2020}.
We use the same k-coefficient Gaussian ordinance nodes (4+4 split between 0-0.95 and 0.95-1.0 \citet{Kataria2013}).

We implement pre-mixed opacities where the high resolution opacities of each species are combined offline and weighted by the mixing ratio of each gas before calculating the k-coefficients across a grid of temperature and pressure points \citep{Showman2009,Amundsen2017}.
\textsc{GGchem} \citep{Woitke2018} is used to calculate the species mixing ratios assuming chemical equilibrium with equilibrium condensation included.
Solar metallicity is assumed using the element ratios presented in \citet{Asplund2009}.
Rayleigh and collision-induced absorption opacities are also added to each band separately during runtime.
Table \ref{tab:line-lists} contains references for the opacities used in the current study.

We use the \textsc{pysynphot} package \citep{pysynphot2013} to calculate the incident stellar flux in each band.
The \textsc{Phoenix} database \citep{Allard2011} was interpolated to the properties of HD 209458 (T$_{\rm eff}$ = 6092 K, R$_{\star}$ = 1.203 R$_{\odot}$ - \textsc{exoplanet.eu}) as the input stellar model.

\subsubsection{Gas phase abundances}

For the current pre-mixed scheme, since the k-table is already weighted by the VMR of each species offline, only the species for the CIA and Rayleigh opacities are required to calculated inside the GCM during runtime.
We use an interpolation from a 2D grid calculated using \textsc{GGchem} including the effects of local equilibrium condensation.
Solar metallicity is assumed using the element ratios presented in \citet{Asplund2009}.
This grid ranges from 691 points in temperature between 100-7000 K in steps of 10 K and 91 pressure points log-spaced between 10$^{-6}$-1000 bar.
Additionally, we also interpolate the local mean molecular weight from the CE calculation for use in the RT scheme.

\section{Initial condition considerations}
\label{sec:IC}

\begin{figure*} 
   \centering
   \includegraphics[width=0.49\textwidth]{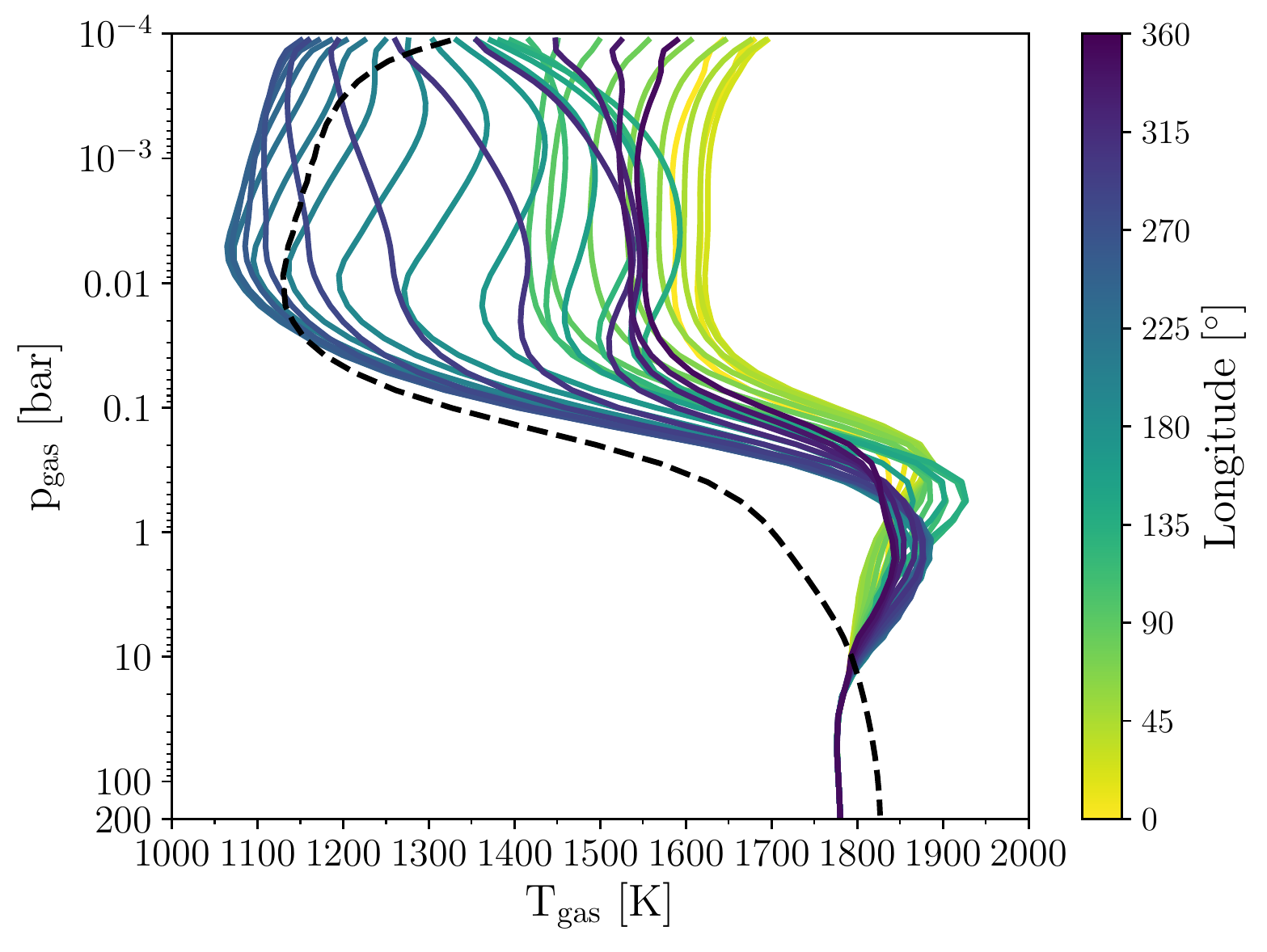}
   \includegraphics[width=0.49\textwidth]{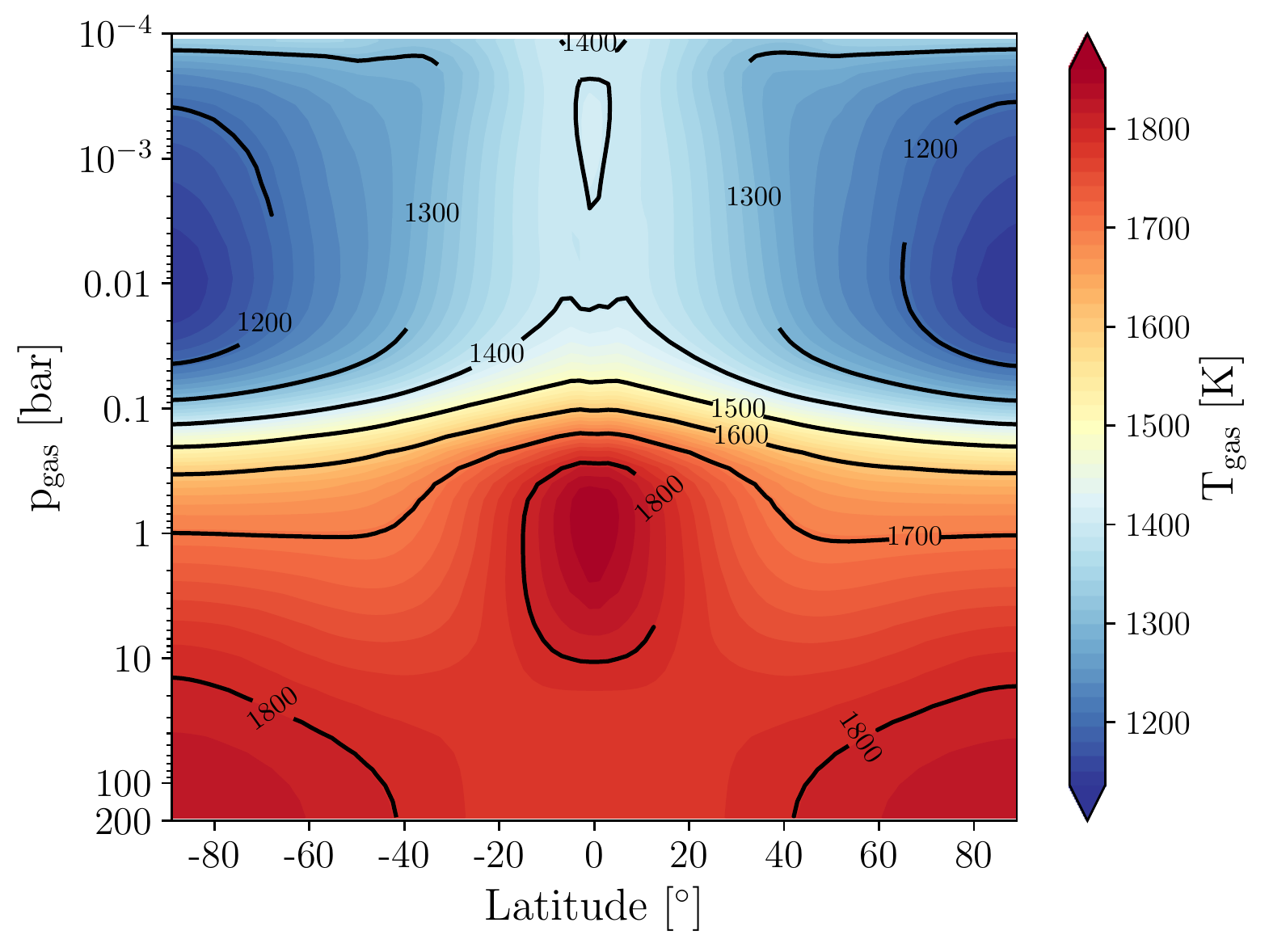}
   \includegraphics[width=0.49\textwidth]{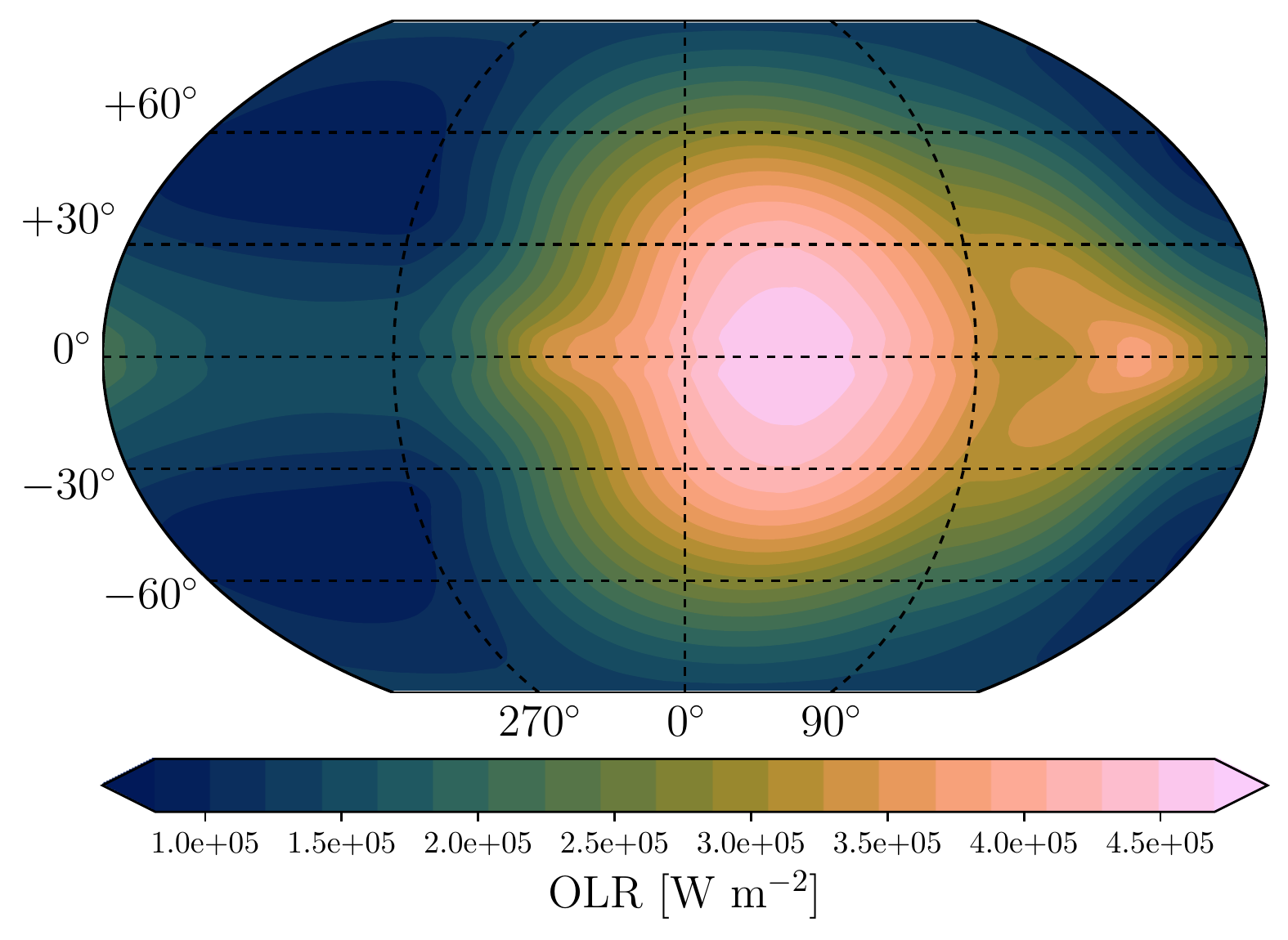}
   \includegraphics[width=0.49\textwidth]{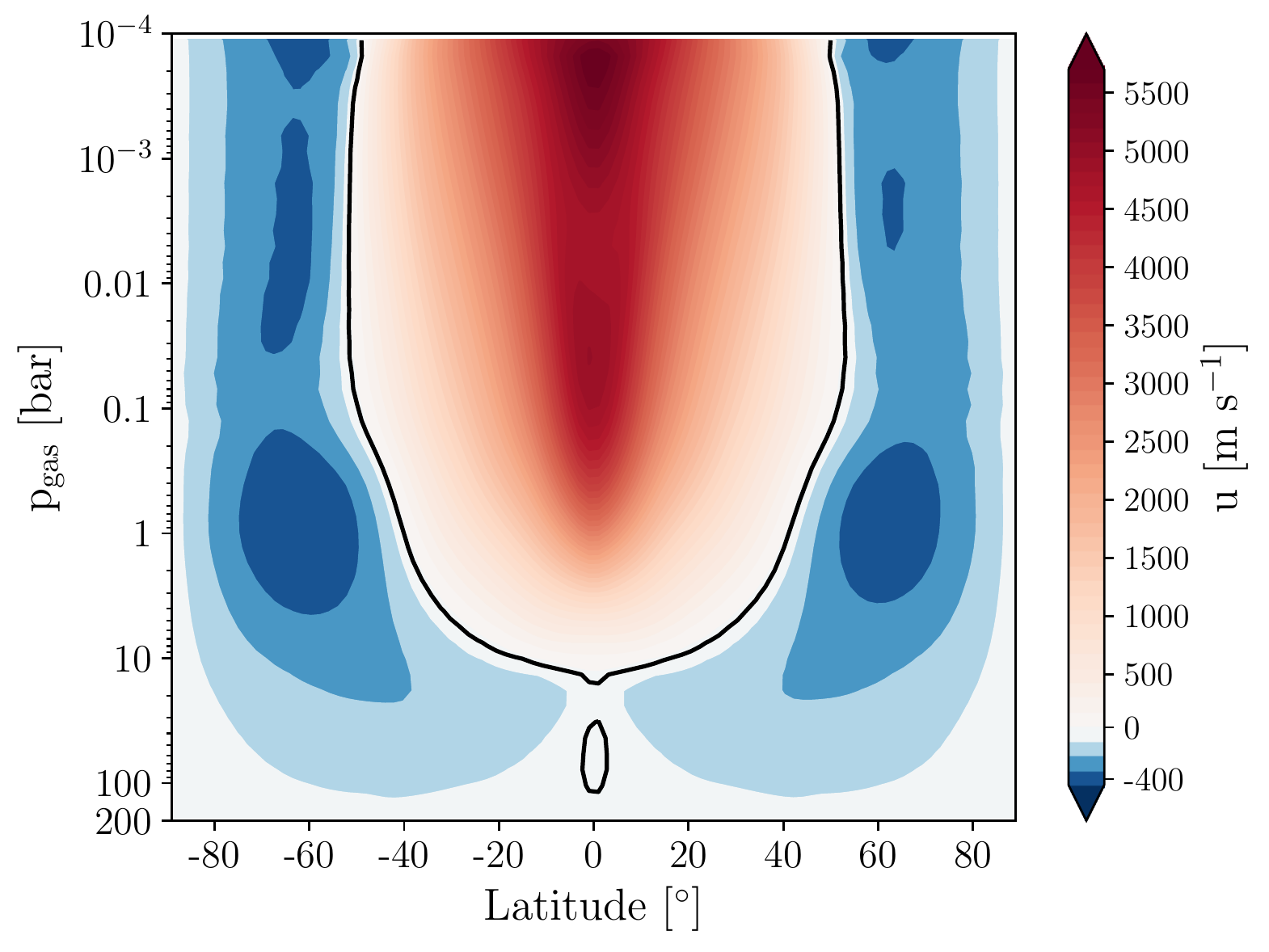}
   \caption{\textsc{Exo-FMS} output of the fiducial \citet{Heng2011b} hot Jupiter simulation parameters.
   Top left: 1D T-p profiles at the equator (solid coloured lines) and polar region (black dashed line).
   Top right: Zonal mean gas temperature.
   Bottom left: Lat-lon map of the OLR flux.
   Bottom right: Zonal mean zonal velocity.}
   \label{fig:Heng}
\end{figure*}

An important consideration, especially for the evolution of the deep layers is the GCM initial conditions (ICs).
We therefore propose a new IC scheme for our three HD 209458b simulations in Sect. \ref{sec:HD209}.

Recently, \citet{Sainsbury-Martinez2019} suggest that the deep atmospheric regions of hot Jupiters tend to evolve to a hot adiabatic profile due to dynamical mixing between the deep and upper layers.
In their models, they found that isothermal ICs gradually evolve towards a deep adiabatic profile over long timescales.
They suggest that a hot adiabatic IC for HJ GCMs should be considered to better capture this long term evolution, as a hotter profile can more quickly cool towards the true adiabatic gradient, compared to a cooler gradient warming up.

\citet{Thorngren2019} published an expression to relate the incident flux and internal temperature in order to fit the observed radii of the hot Jupiter population.
This greatly affects the thermal structure of the deep atmosphere, with larger T$_{\rm int}$ increasing the steepness of the T-p structure.

Bearing these studies in mind, we initialise each 1D GCM column in our HD 209458b tests using the \citet{Parmentier2014, Parmentier2015} analytical T-p expressions at the sub-stellar point ($\mu_{\star}$ = 1) with their adiabatic correction scheme.
Using the sub-stellar point profile ensures that the adiabatic region will be slightly too hot initially for most of the atmosphere, allowing the suggestion of \citet{Sainsbury-Martinez2019} to be implemented in a simple manner.
The T$_{\rm int}$ value (here 571 K for HD 209458b) is calculated using the \citet{Thorngren2019} expression of the internal temperatures of the hot Jupiter population as a boundary condition for the IC scheme as well as setting the internal flux for the RT scheme.

\section{Semi-Grey RT results}
\label{sec:semi-grey}

In this section, we present results using the \textsc{Exo-FMS} semi-grey RT scheme, benchmarking to the \citet{Heng2011b} and \citet{Rauscher2012} studies that both used a semi-grey RT approach.

\subsection{Heng et al. (2011) benchmark}
\label{sec:Heng}

In \citet{Heng2011b} simulations were performed with the FMS spectral dynamical core including a grey gas RT scheme.
Since their experiments used a similar framework to our current study, \citet{Heng2011b} is the most comparable simulation to the current study.
We perform the fiducial hot Jupiter model with the parameters found in \citet[][Table. 1]{Heng2011b}.
We alter the grey RT in Sect. \ref{sec:grey_RT} to conform to the same scheme in \citet{Heng2011b}.
An isothermal initial condition at 1824 K is used with T$_{\rm int}$ = 0 K internal temperature.
The simulation is run for 3600 (Earth) days, with the final 100 day average presented as the results in this study.

Figure \ref{fig:Heng} presents our benchmarking results to \citet{Heng2011b}.
The zonal mean wind and temperature structures closely match the results of \citet{Heng2011b}.
The wind has a peak value of 5500 m s$^{-1}$ occurring near 0.1 bar in both cases, and the jet widths are similar.
The temperature hot spot is 1800 K occurring near 1 bar in both cases, and the horizontal pattern of OLR, which is an important signature of wave dynamics, agrees closely with \citet{Heng2011b}.
Both simulations show prominent Rossby wave cold cyclones straddling the equator to the west of the substellar point.
This structure is an important indicator of the effect of the jet on the wave dynamics \citep{Hammond2018}.
The GCM in \citet{Heng2011b} used the spectral dynamical core and here the cube sphere, finite volume version.
This is an indication the Exo-FMS dynamical core is producing comparable flow patterns compared to contemporary models.

The agreement in jet structure is notable in view of the very different dynamical cores used in the two simulations and demonstrates that the use of the cube sphere grid has not distorted the relevant angular momentum transport processes, since the jet speed is sensitive to angular momentum transport \citep{Mendonca2020}.

Angular momentum transport plays a crucial role in the generation of super-rotation, and problems with angular momentum conservation have been implicated as a cause of intermodal differences in reproducing the superstation of Venus \citep{Lee2012b}.
The FMS spectral dynamical core has been found to conserve angular momentum in Venus test cases to within 2$\%$ per century \citep{Lee2012b}.
However, the momentum conservation properties for the hot Jupiter class of exoplanets is different to that of Venus.
\citet{Polichtchouk2014} discuss the differences in momentum conservation between GCM models in a hot Jupiter context, and included the MITgcm cubed-sphere dynamical core in their analysis.
However, the FMS cube-sphere (or spectral) dynamical cores were not compared in \citet{Polichtchouk2014}, which has different numerical schemes to MITgcm.
The good agreement between our FMS cube-sphere simulations and the \citet{Heng2011b} spectral benchmark is a reassuring indication that the cube sphere dynamical core can adequately reproduce the necessary angular momentum transport properties, at least in the context of tide-locked circulations.
However, this is not a rigorous test of the conservation properties of the current model.
With our additions to Exo-FMS in this paper, a more substantive intercomparison, similar to that of \citet{Polichtchouk2014}, can be performed in hot Jupiter context in future dedicated investigations.

\subsection{Rauscher \& Menou (2012) benchmark}
\label{sec:Rauscher}

\begin{figure*} 
   \centering
   \includegraphics[width=0.49\textwidth]{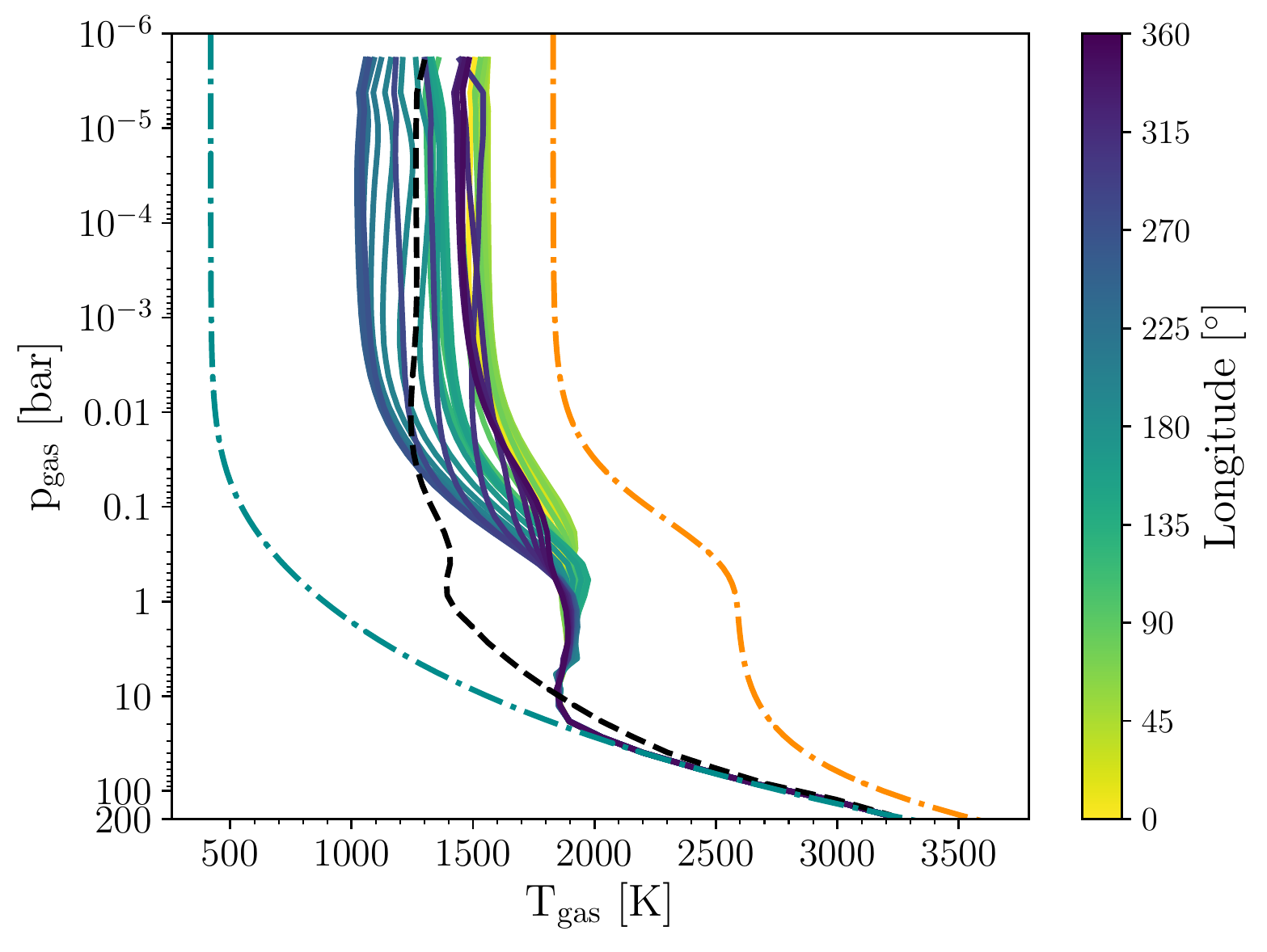}
   \includegraphics[width=0.49\textwidth]{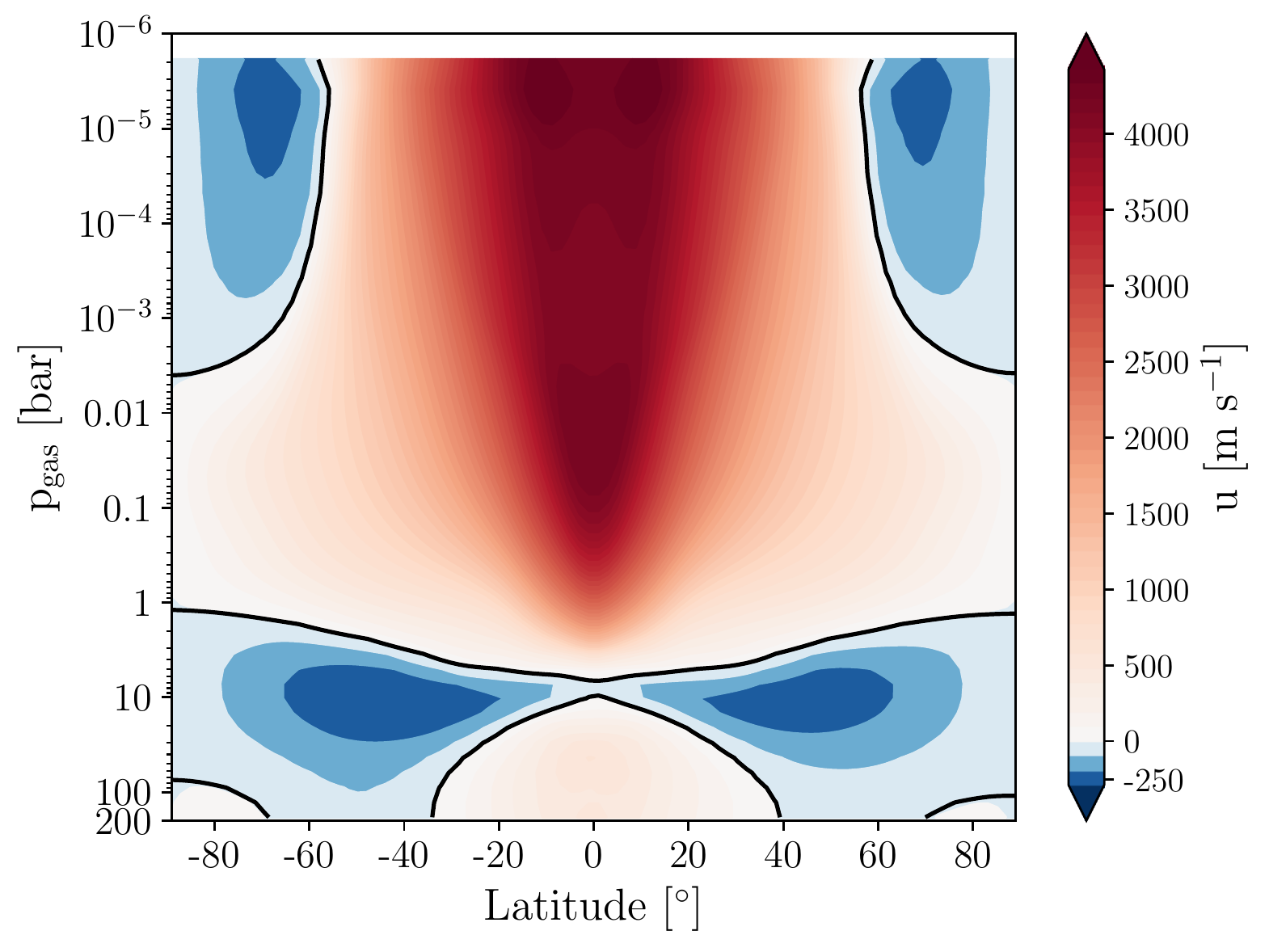}
   \includegraphics[width=0.49\textwidth]{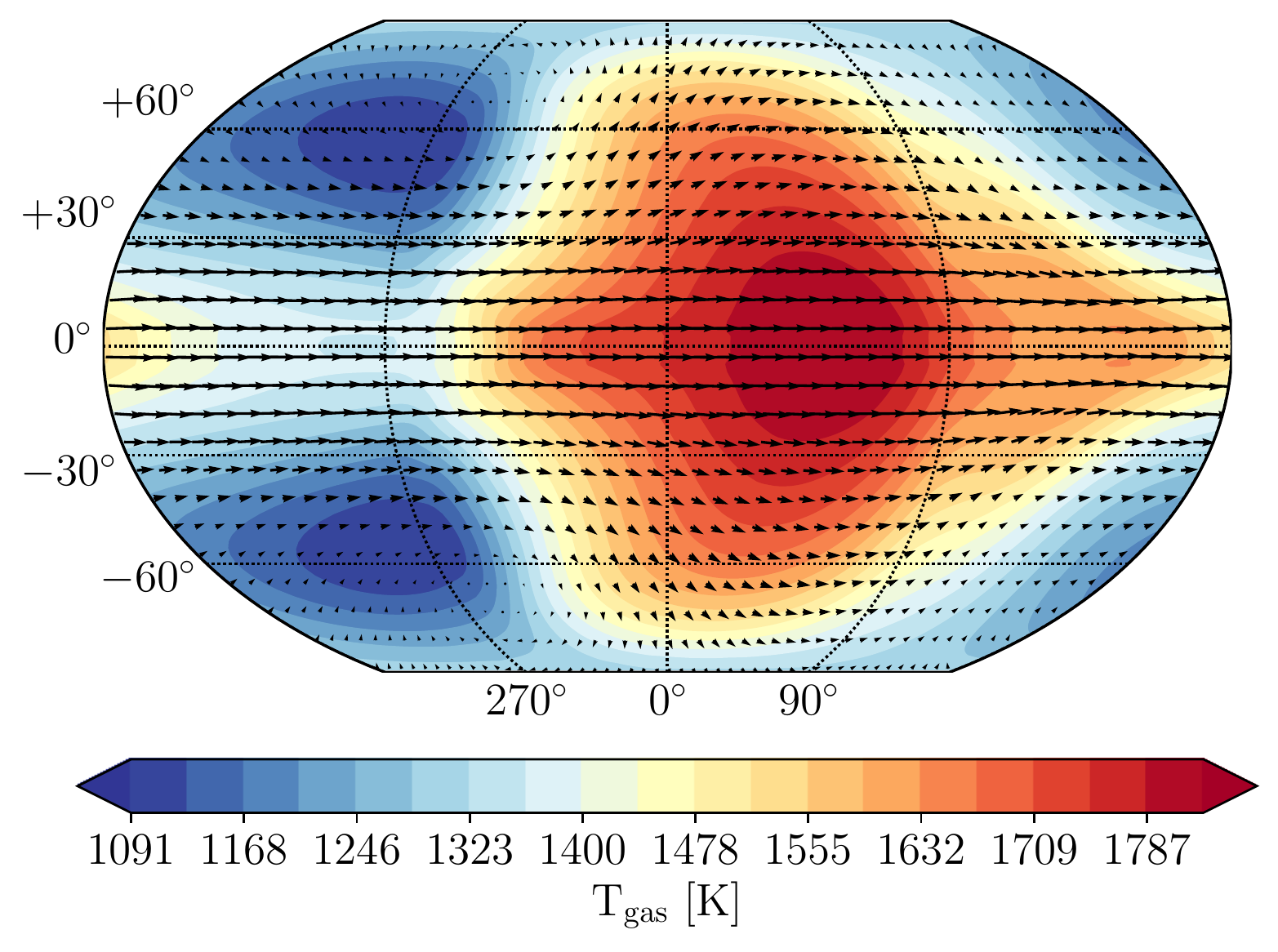}
   \includegraphics[width=0.49\textwidth]{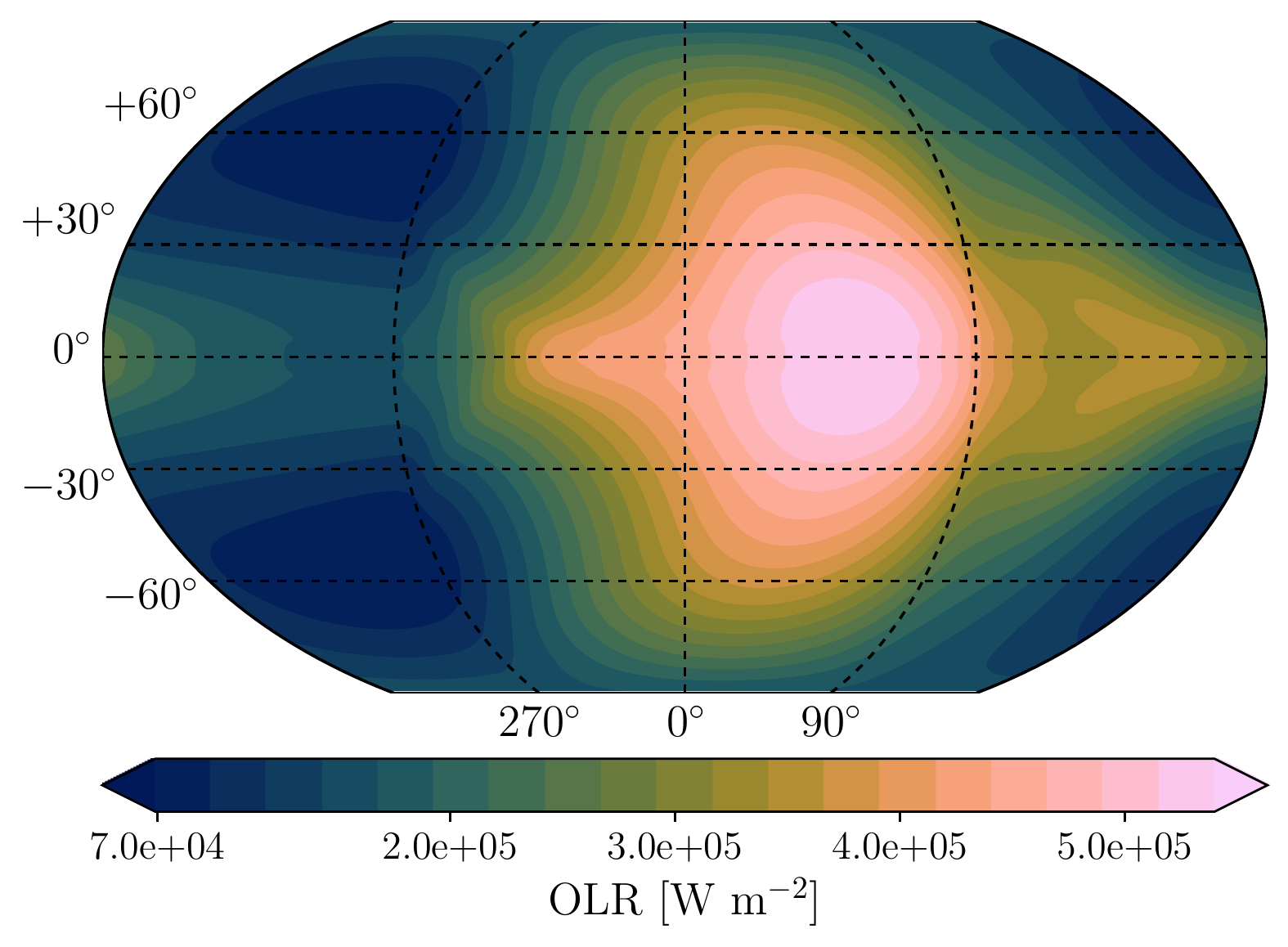}
   \caption{\textsc{Exo-FMS} output of the hot Jupiter simulation presented in \citet{Rauscher2012}.
   Top left: 1D equatorial T-p profiles (solid colour lines) and polar region (black dashed line).
   The dash-dot lines show the sub-stellar (orange) and self-luminous (cyan) radiative-equilibrium solutions from \citet{Guillot2010}.
   Top right: zonal mean zonal velocity.
   Bottom left: lat-lon map of gas temperature with velocity vectors (black arrows) near the photosphere ($\approx$ 51 mbar).
   Bottom right: lat-lon map of the OLR flux.}
   \label{fig:Rauscher}
\end{figure*}

In \citet{Rauscher2012} several double-grey RT models were performed to examine the effects of magnetic drag on the dynamical properties of hot Jupiter exoplanets.
The model used in \citet{Rauscher2012}, like that in \citet{Heng2011b}, employs a spectral dynamical core, but derives from an independently developed model with somewhat different formulation of the primitive equations, in long use for Earth climate studies.
We perform a similar simulation to their fiducial model present in Table. 1 of \citet{Rauscher2012}.
\citet{Rauscher2012} include a flux limited diffusion scheme for longwave radiation at high optical depths.
As in \citet{Rauscher2012} we initialise the model with Milne's T-p profile solution for self-luminous objects \citep[e.g.][]{Guillot2010, Heng2014b} with T$_{\rm int}$ = 500 K.

In Fig. \ref{fig:Rauscher} we present the results of our \textsc{Exo-FMS} run using the \citet{Rauscher2012} simulation parameters.
Results are taken after 3600 days of simulation, with the final 100 day average used as the presented results.
The qualitative features of the atmospheric state agree with our simulation, in that there is a super-rotating equatorial jet which decays with depth in the atmosphere, a hot-spot in the OLR which is shifted eastward relative to the sub-stellar point, and a pronounced dayside-nightside temperature difference in the upper atmosphere (as seen in the profiles).
However, there are a number of significant quantitative differences.
The jet maximum in \citet{Rauscher2012} is about 2000 m s$^{-1}$ stronger, and the jet penetrates deeper into the atmosphere (about 10 bars vs. about 3 bars in our case).
The nightside equatorial temperature profiles agree well between the two simulations, but the dayside temperature in \citet{Rauscher2012} has a peak of 2200 K at 0.5 bars, whereas our simulation is monotonic in the upper atmosphere and several hundred K cooler at the 0.5 bar level, however, the dayside temperatures in the far upper atmosphere agree reasonably well between the simulations.
Another quantitative difference is that the hot-spot shift is approximately 25 degrees in \citet{Rauscher2012}, vs. 50 degrees in our simulation.

Because Exo-FMS simulations check out well against the \citet{Heng2011b} simulations, which like \citet{Rauscher2012} employed a spectral dynamical core, we think it unlikely that the quantitative differences arise from the dynamical core.
Although we have set the grey parameters in our radiative scheme to mimic \citet{Rauscher2012} as closely as possible, the differences in the formulation of the radiation scheme make an exact match impossible.
This disparity most likely stems from the differences in the RT scheme used between the models, with \citet{Rauscher2012} switching to a diffusion approximation at high IR optical depths, whereas we use a two-stream approach with a linear in tau term (Sect. \ref{sec:lw}).
Differences in the cooling/heating rates leads a variance in the dynamical forcing between the models, however, it is difficult in practice to diagnose the exact reason for the different results.
We take the discrepancies as an indication that even within the scope of semi-grey radiation, differences in the formulation can lead to significantly different results.

\section{HD 209458b-like simulations}
\label{sec:HD209}

In this section we perform HD 209458b-like simulations, similar to those in \citet{Showman2009}.
The simulation in \citet{Showman2009} employs the MITgcm dynamical core, which, like Exo-FMS, uses a cube-sphere grid with finite-volume numerics, though it is coded independently from Exo-FMS and involves somewhat different numerical algorithms.
This simulation is then run using the semi-grey, non-grey picket fence and a correlated-k model.
The difference between each model are then compared and examined on how each RT scheme affected the end results.
The parameters used in each experiment can be found in Appendix \ref{app:GCM_parameters}.

\subsection{Semi-grey RT results}
\label{sec:semi-grey2}

\begin{figure*} 
   \centering
   \includegraphics[width=0.49\textwidth]{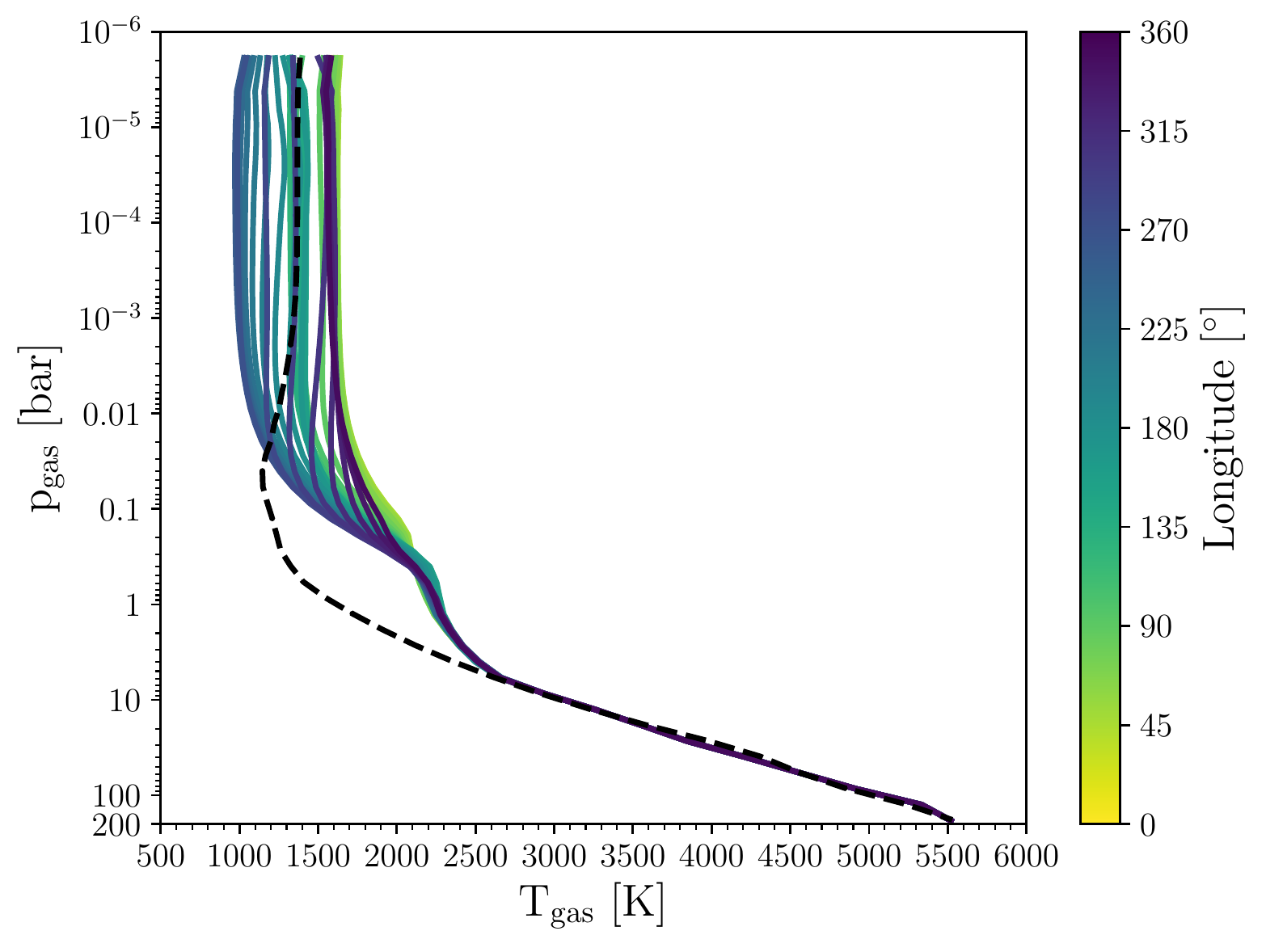}
   \includegraphics[width=0.49\textwidth]{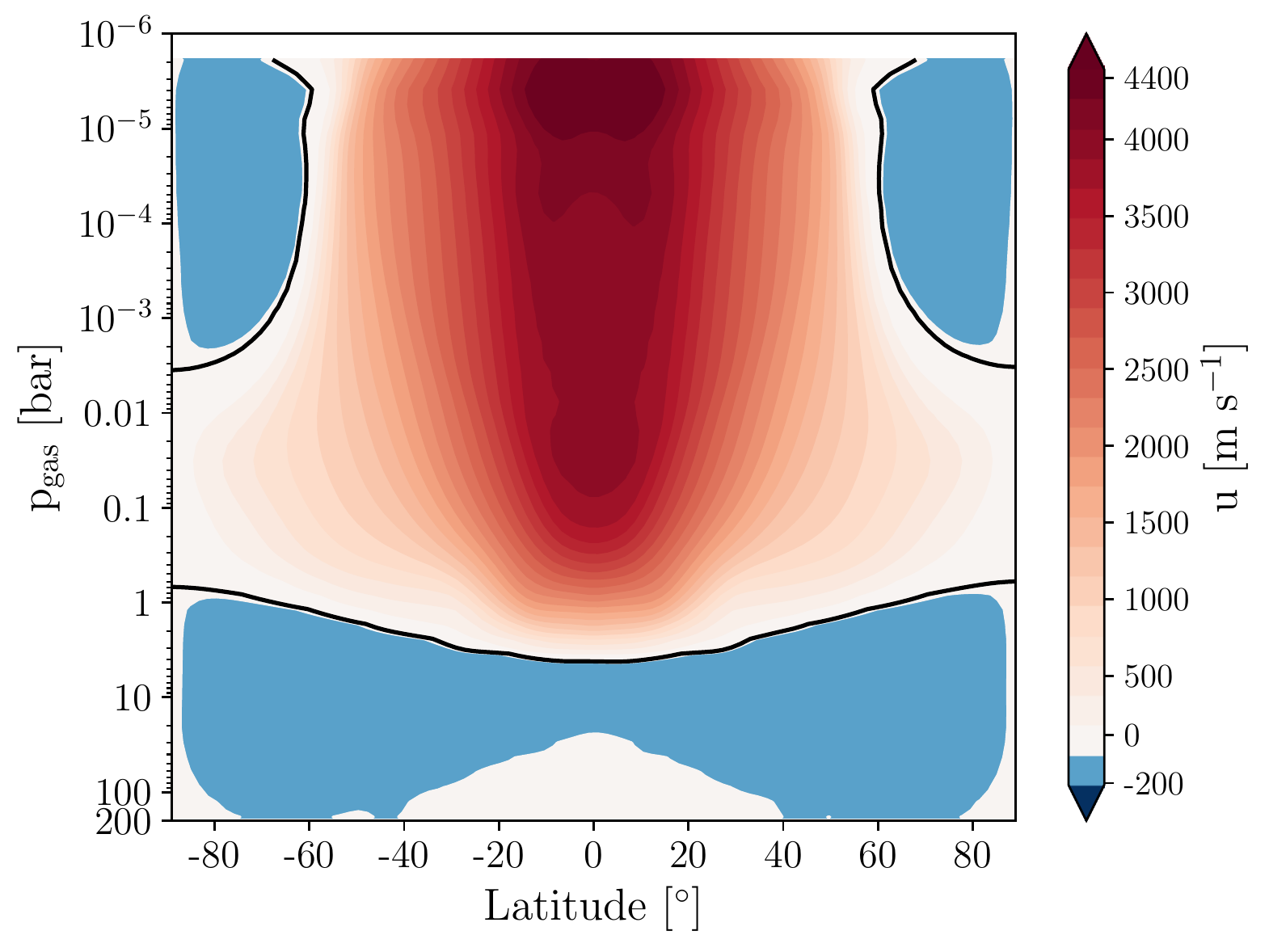}
   \includegraphics[width=0.49\textwidth]{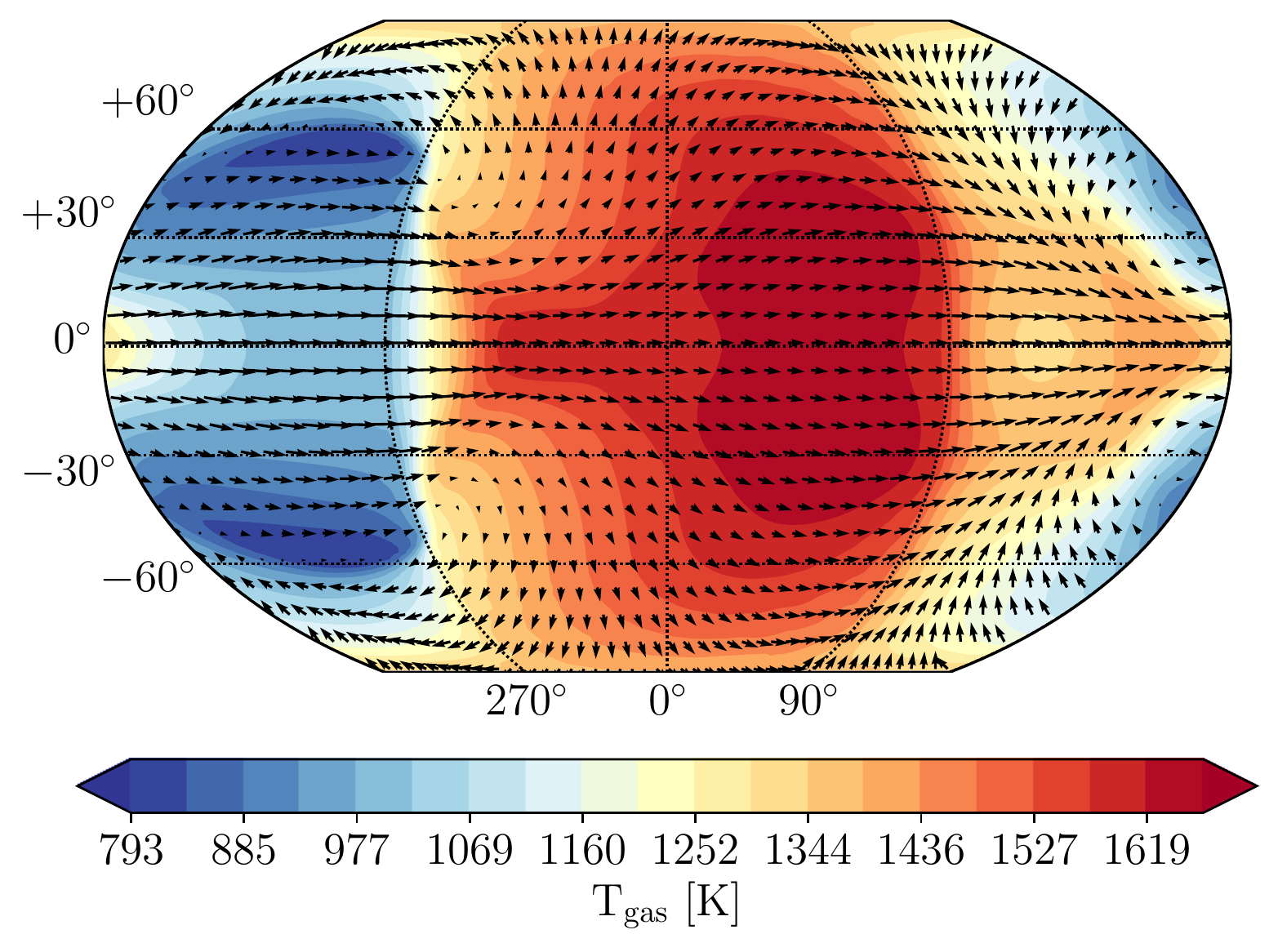}
   \includegraphics[width=0.49\textwidth]{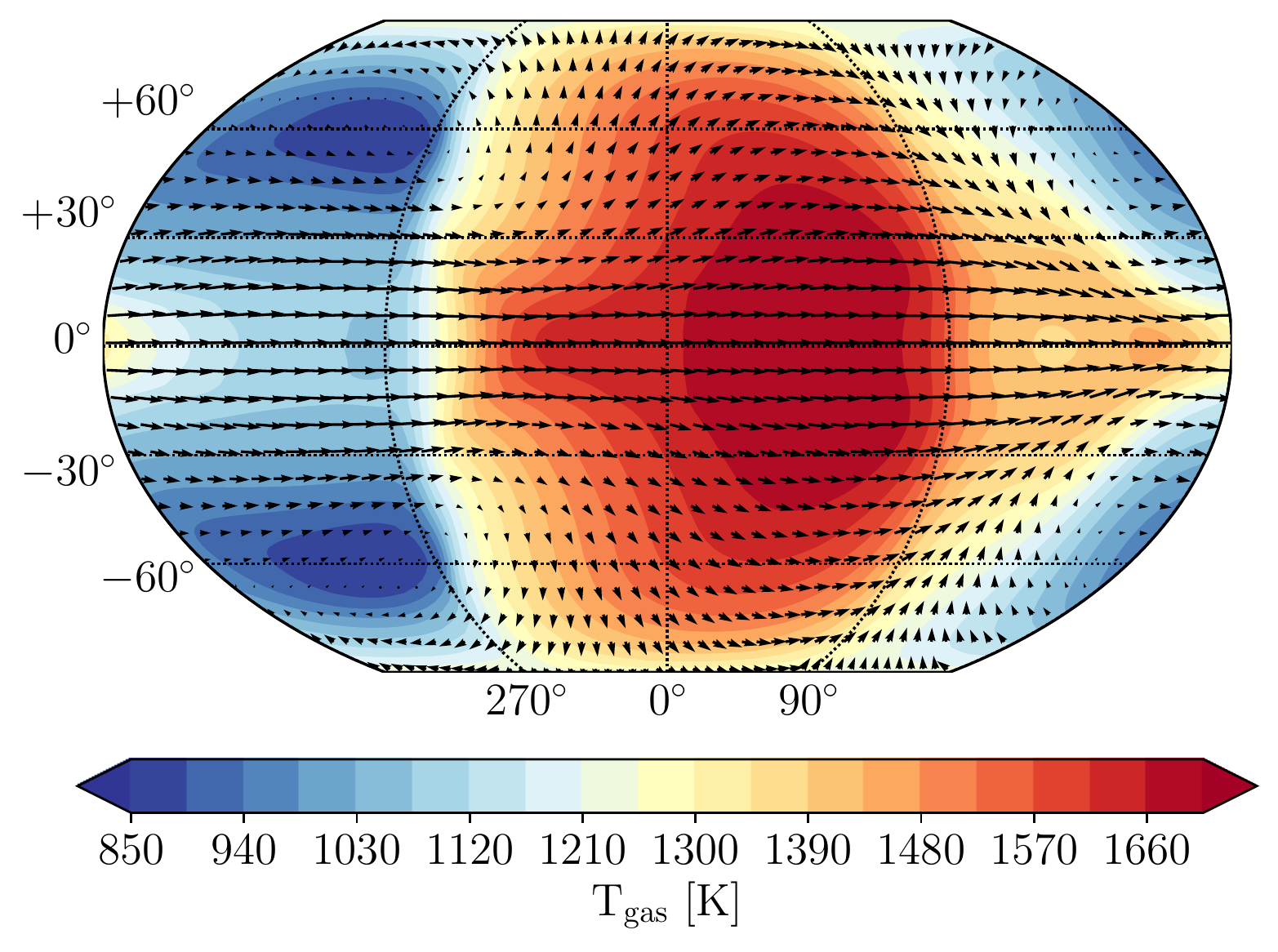}
   \includegraphics[width=0.49\textwidth]{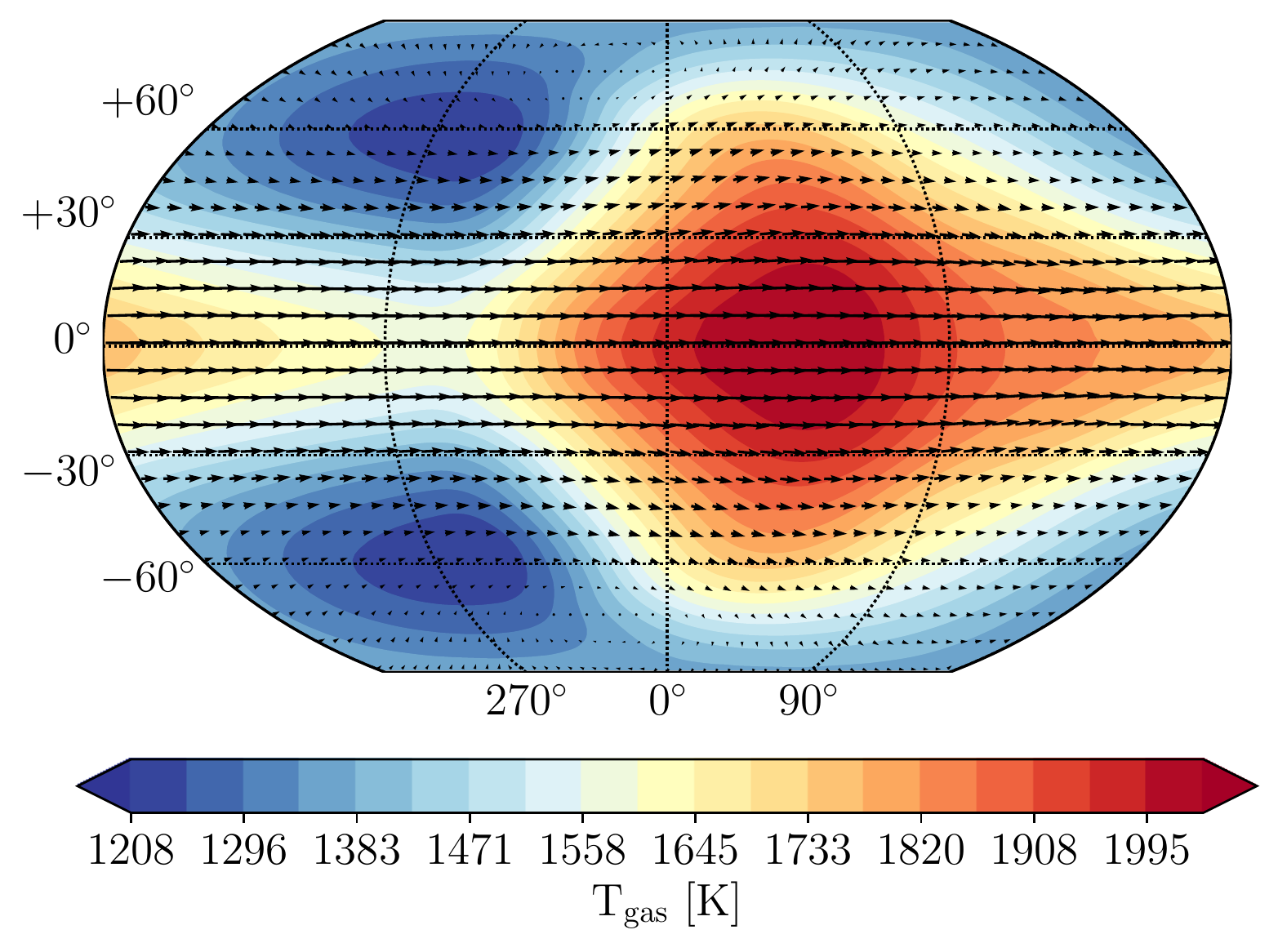}
   \includegraphics[width=0.49\textwidth]{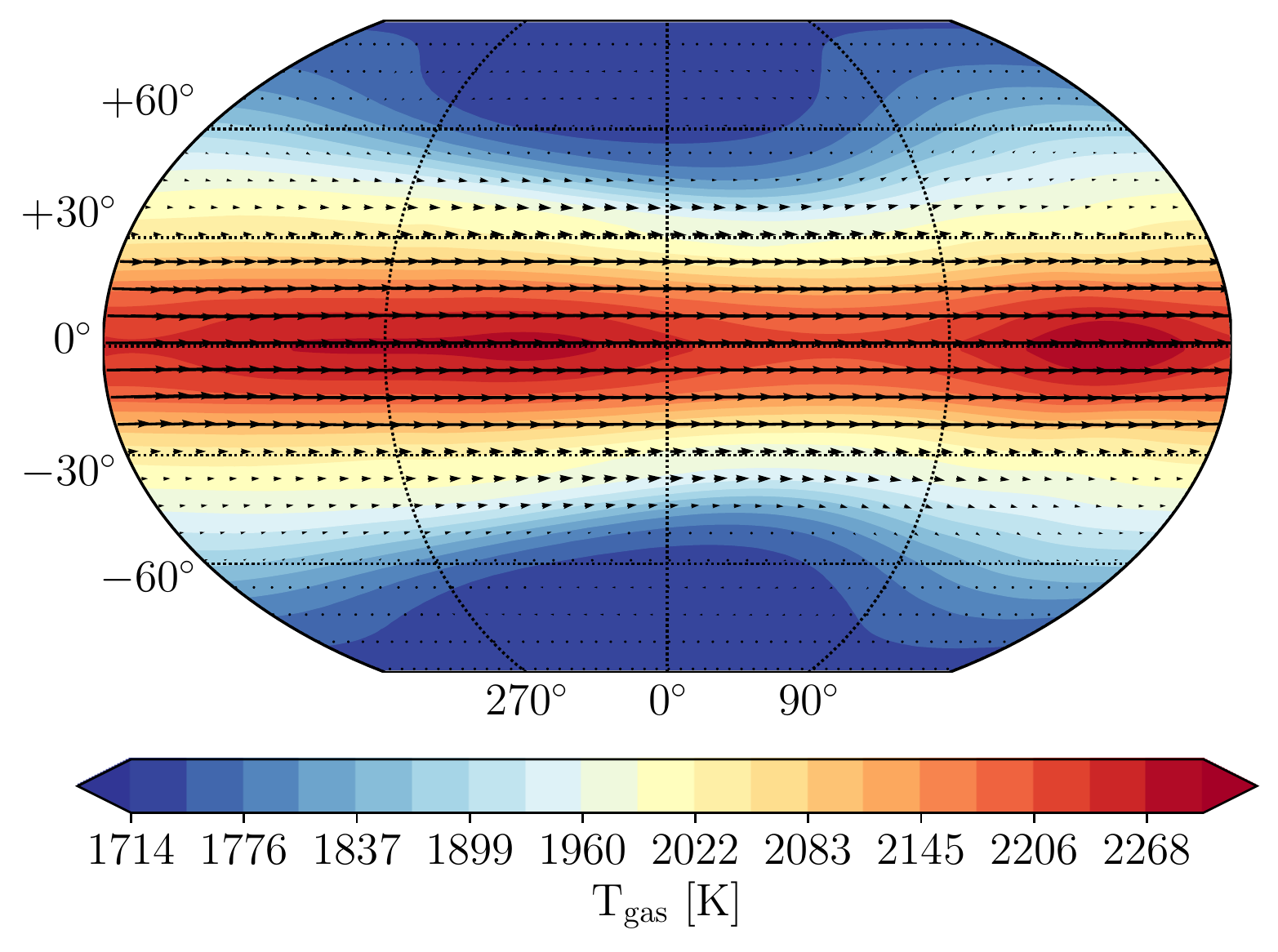}
   \caption{\textsc{Exo-FMS} output of a HD 209458b-like experiment using the semi-grey RT scheme.
   Top left: 1D equatorial T-p profiles (solid colour lines) and polar region (black dashed line).
   Top right: zonal mean zonal velocity.
   Middle left: lat-lon map of gas temperature at $\approx$1 mbar.
   Middle right: lat-lon map of gas temperature at $\approx$10 mbar.
   Bottom left: lat-lon map of gas temperature at $\approx$0.1 bar.
   Bottom right: lat-lon map of gas temperature at $\approx$1 bar.
   Velocity vectors are shown as back arrows.}
   \label{fig:HD209_sg}
\end{figure*}

In this simulation we use the semi-grey RT scheme for the HD 209458b-like simulation.
For the semi-grey opacities, we use the \citet{Guillot2010} values of $\kappa_{IR}$ = 10$^{-3}$ [m$^{2}$ kg$^{-1}$] and $\kappa_{V}$ = 6 $\times$ 10$^{-4}$ $\times$ $\sqrt{T_{\rm irr}/2000}$ = 6.14 $\times$ 10$^{-4}$ [m$^{2}$ kg$^{-1}$].
\citet{Guillot2010} derived semi-grey analytic T-p profiles under the assumption of radiative-equilibrium.
In \citet{Guillot2010} these grey opacity values were found to best fit the T-p profile of more sophisticated models of HD 209458b's atmosphere.
The simulation is initialised according to scheme outlined in Sect. \ref{sec:IC}.
The simulation is run for 3600 days, with the final 100 day average taken as the results.
Figure \ref{fig:HD209_sg} presents the equatorial T-p profiles, zonal-mean velocity and lat-lon pressure level temperature maps.

\subsection{Non-grey picket fence RT results}
\label{sec:non-grey}

\begin{figure*} 
   \centering
   \includegraphics[width=0.49\textwidth]{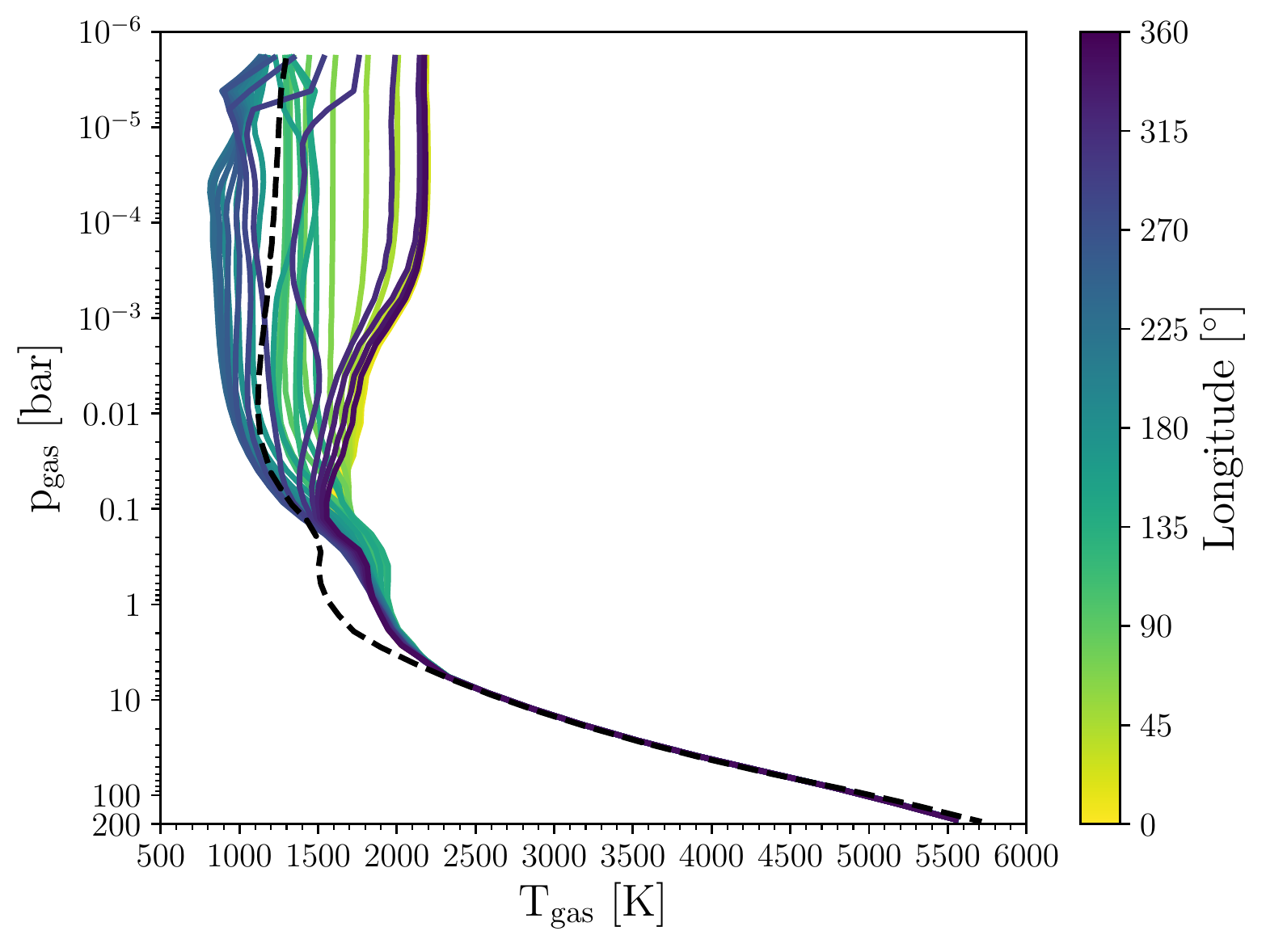}
   \includegraphics[width=0.49\textwidth]{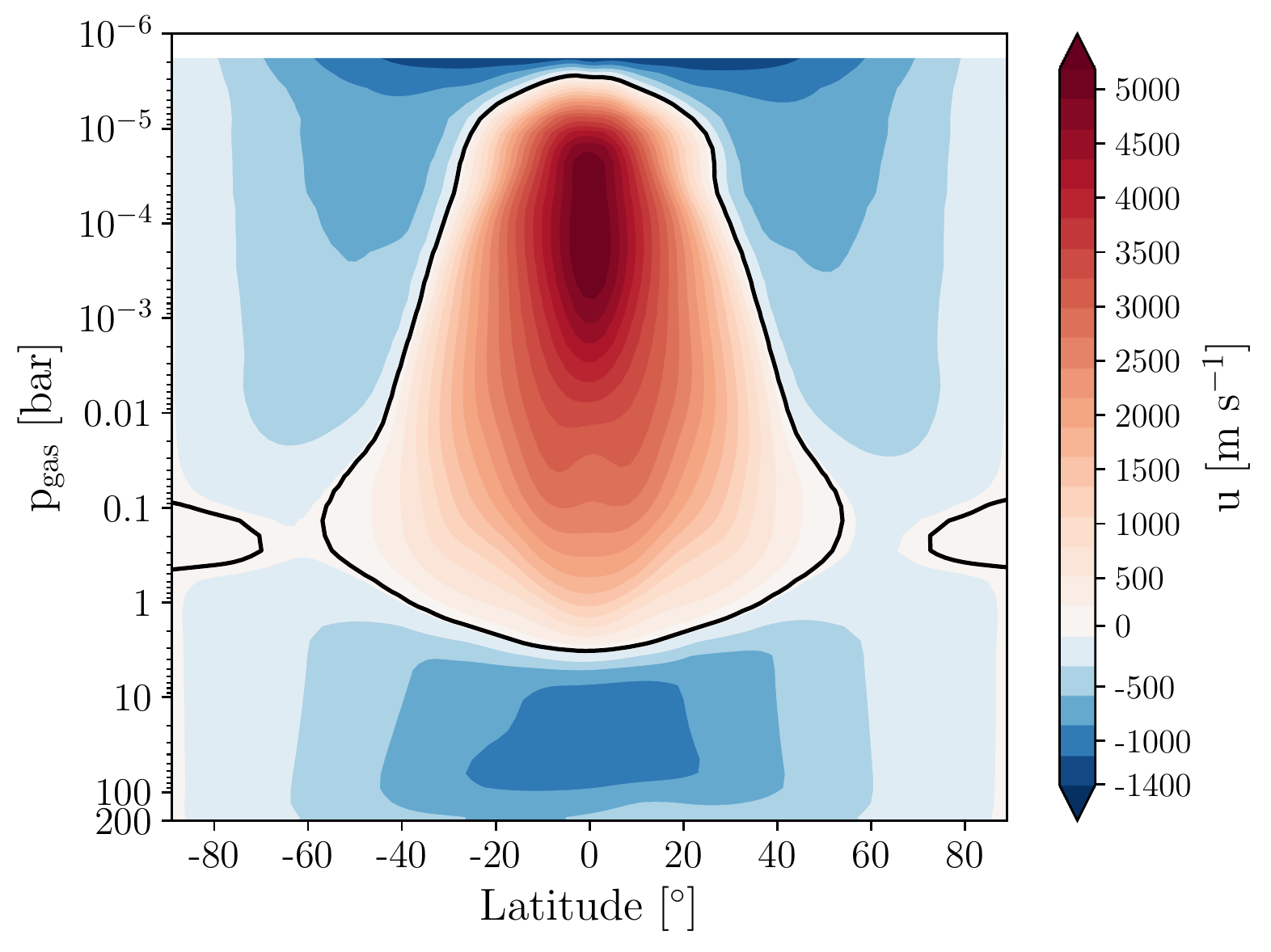}
   \includegraphics[width=0.49\textwidth]{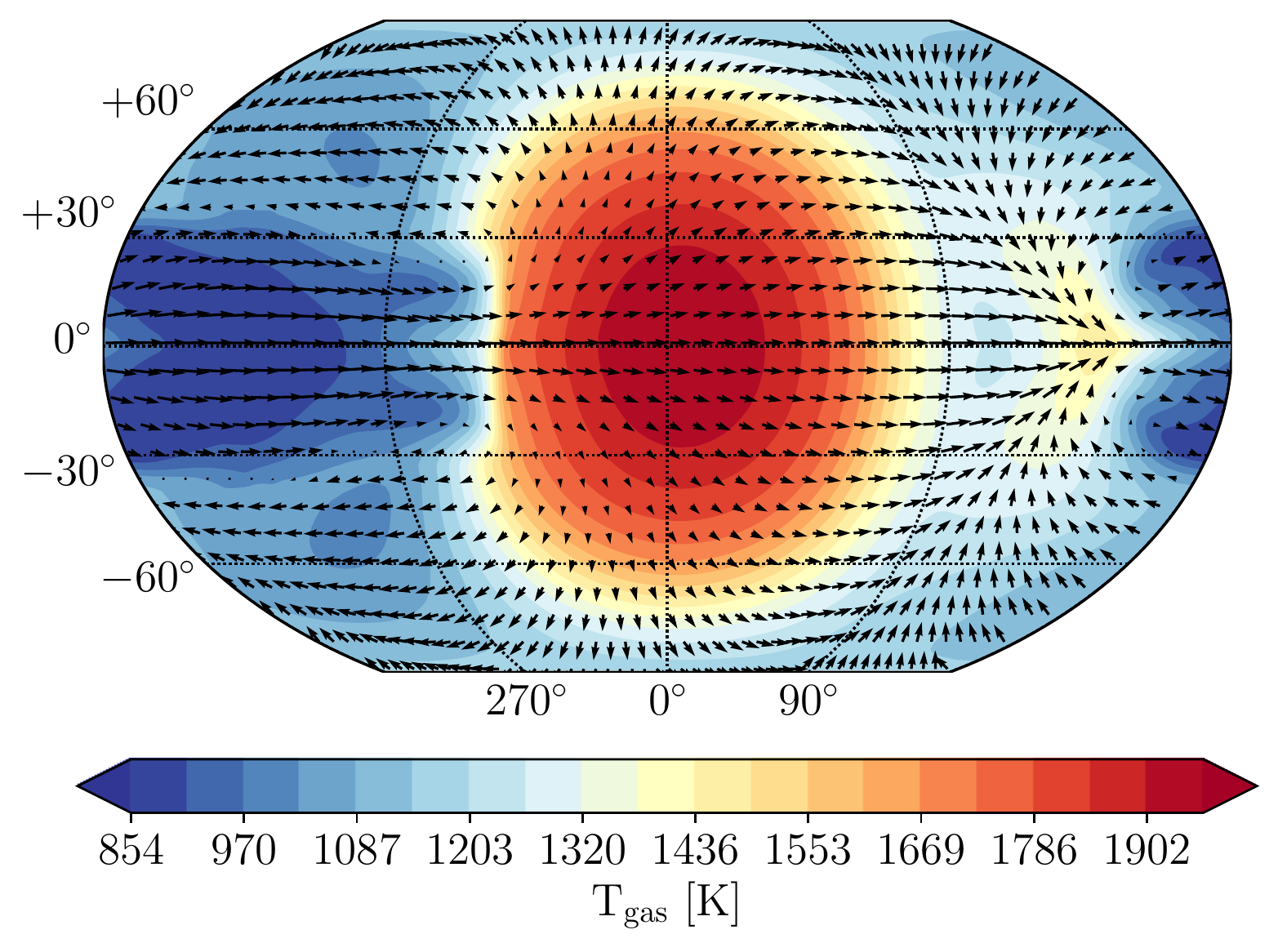}
   \includegraphics[width=0.49\textwidth]{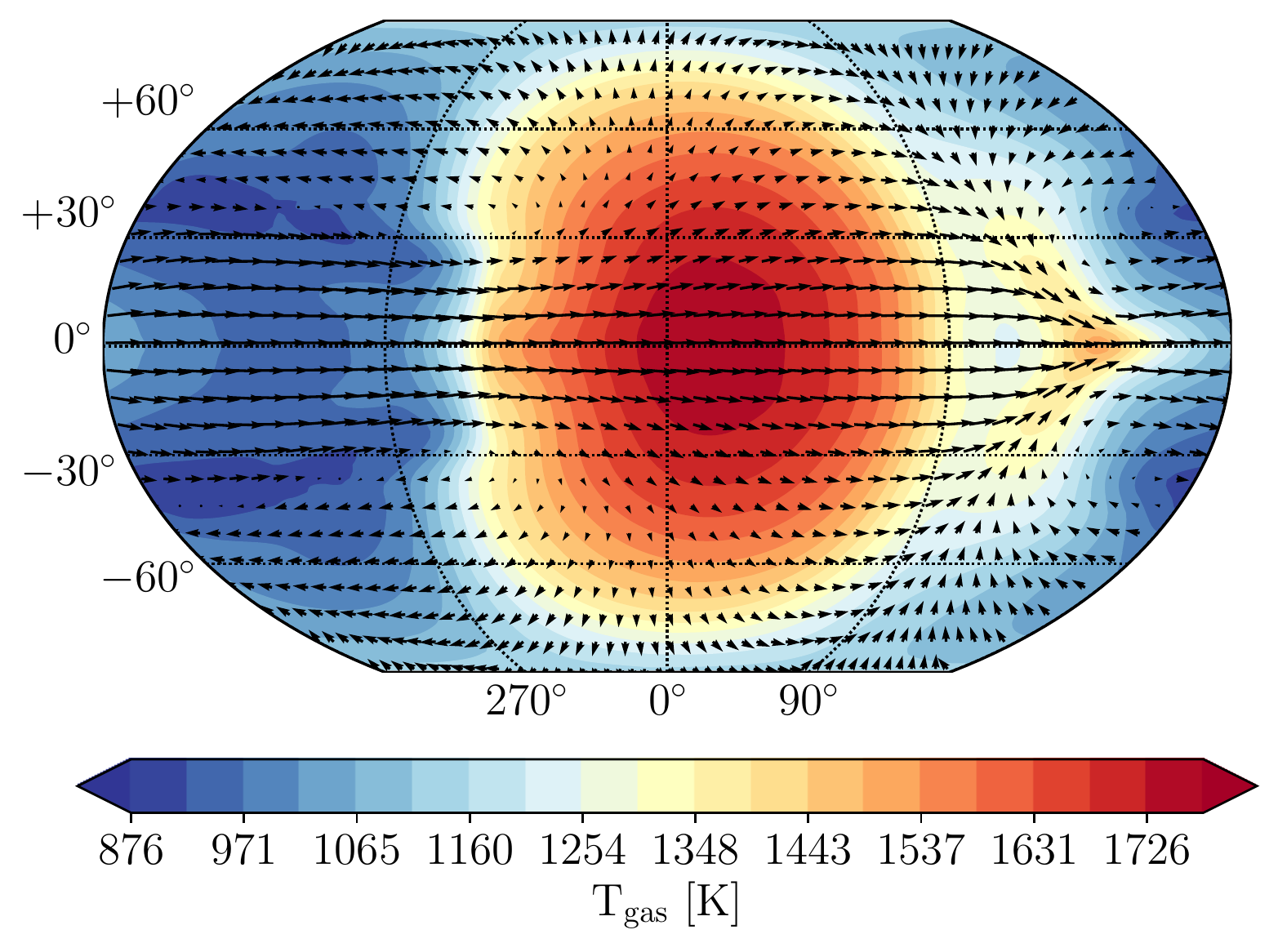}
   \includegraphics[width=0.49\textwidth]{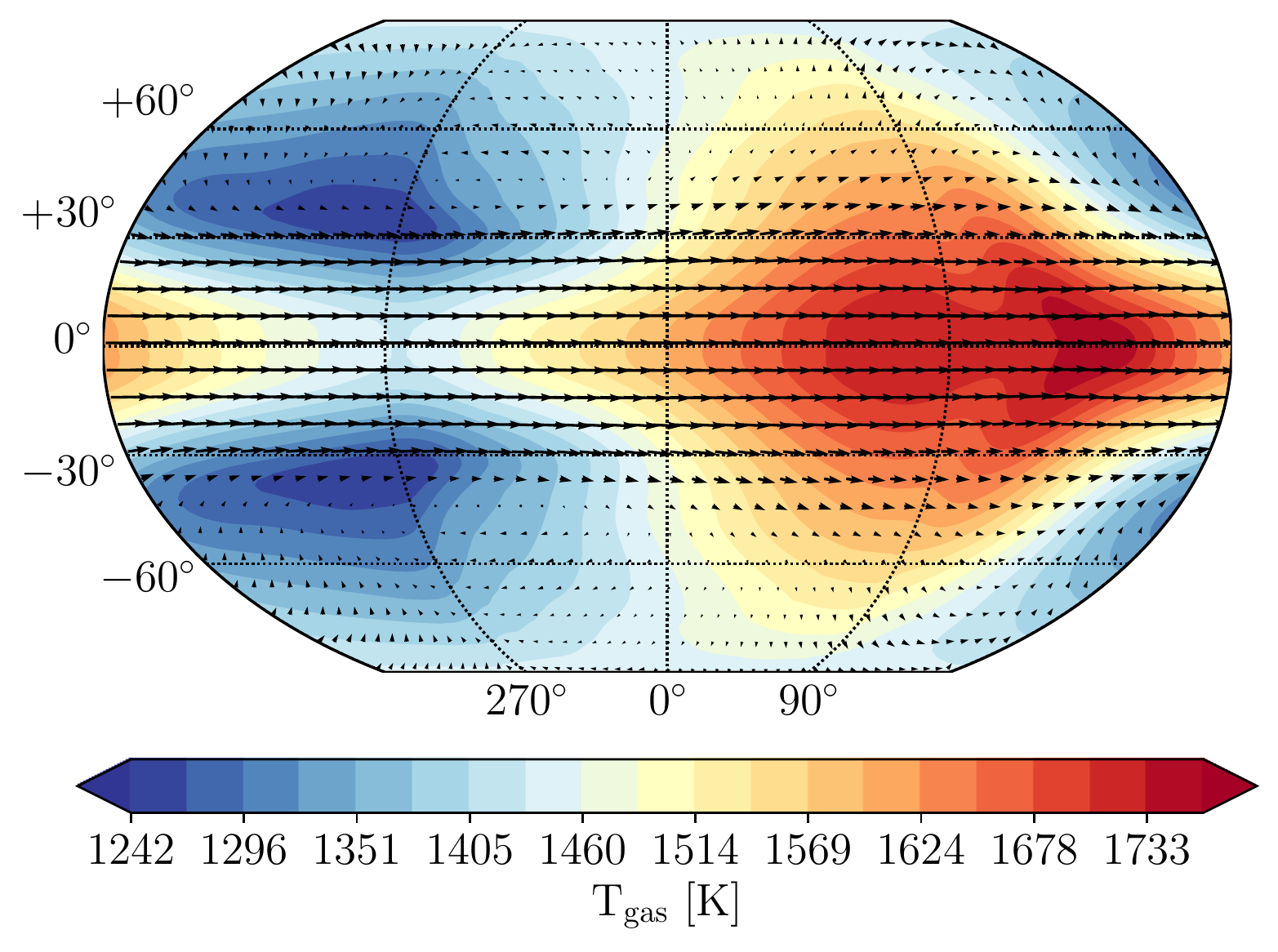}
   \includegraphics[width=0.49\textwidth]{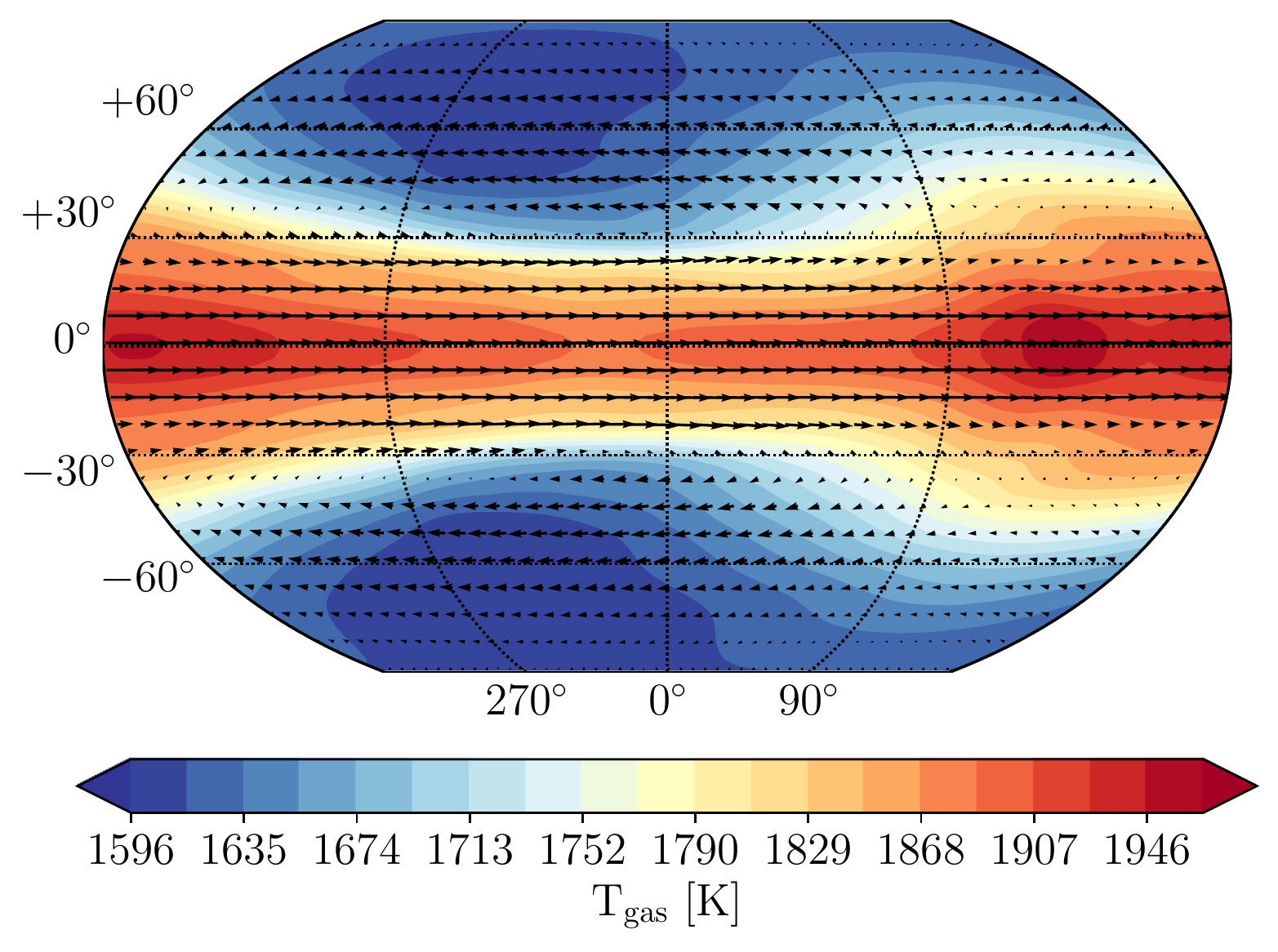}
   \caption{\textsc{Exo-FMS} output of a HD 209458b-like experiment using the non-grey picket fence scheme.
   Top left: 1D equatorial T-p profiles (solid colour lines) and polar region (black dashed line).
   Top right: zonal mean zonal velocity.
   Middle left: lat-lon map of gas temperature at $\approx$1 mbar.
   Middle right: lat-lon map of gas temperature at $\approx$10 mbar.
   Bottom left: lat-lon map of gas temperature at $\approx$0.1 bar.
   Bottom right: lat-lon map of gas temperature at $\approx$1 bar.
   Velocity vectors are shown as back arrows.}
   \label{fig:HD209_non}
\end{figure*}

We present the results using the picket fence RT modifications from Sect. \ref{sec:non-grey_RT}.
The simulation is initialised according to scheme outlined in Sect. \ref{sec:IC}.
The simulation is run for 3600 days, with the final 100 day average taken as the results.
Figure \ref{fig:HD209_non} presents the results of this simulation.

\subsection{Real gas RT results}
\label{sec:spectral}

\begin{figure*} 
   \centering
   \includegraphics[width=0.49\textwidth]{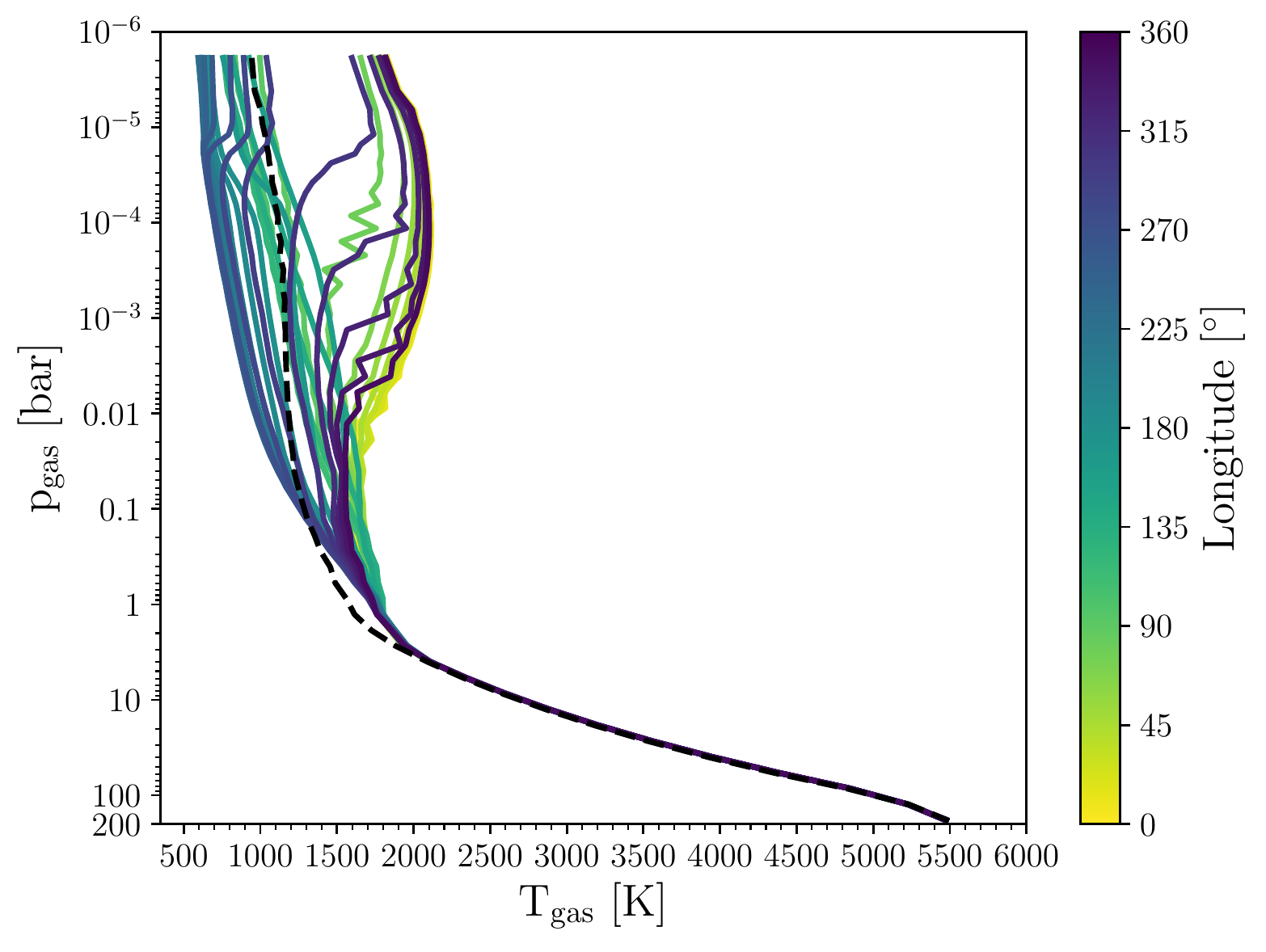}
   \includegraphics[width=0.49\textwidth]{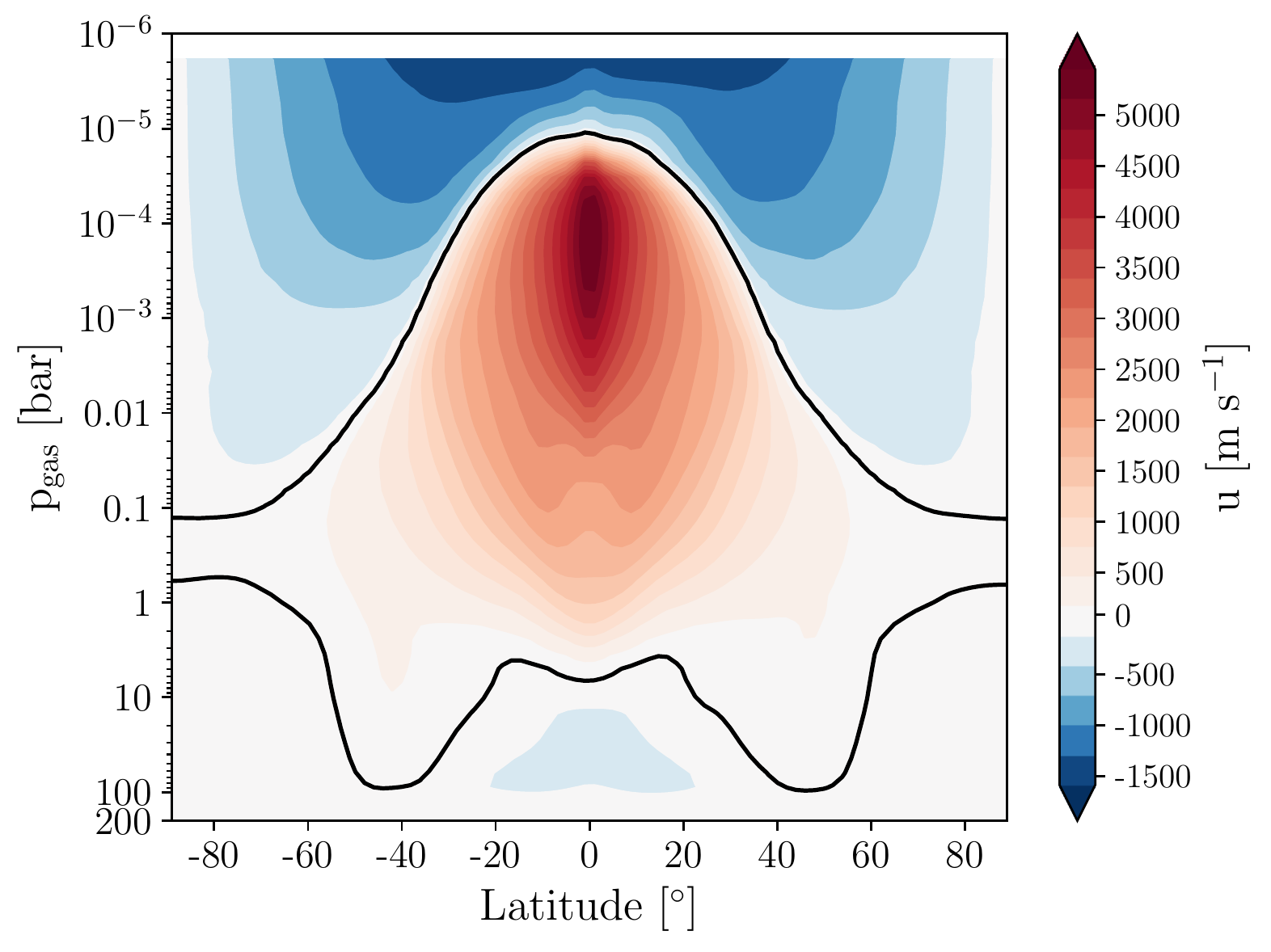}
   \includegraphics[width=0.49\textwidth]{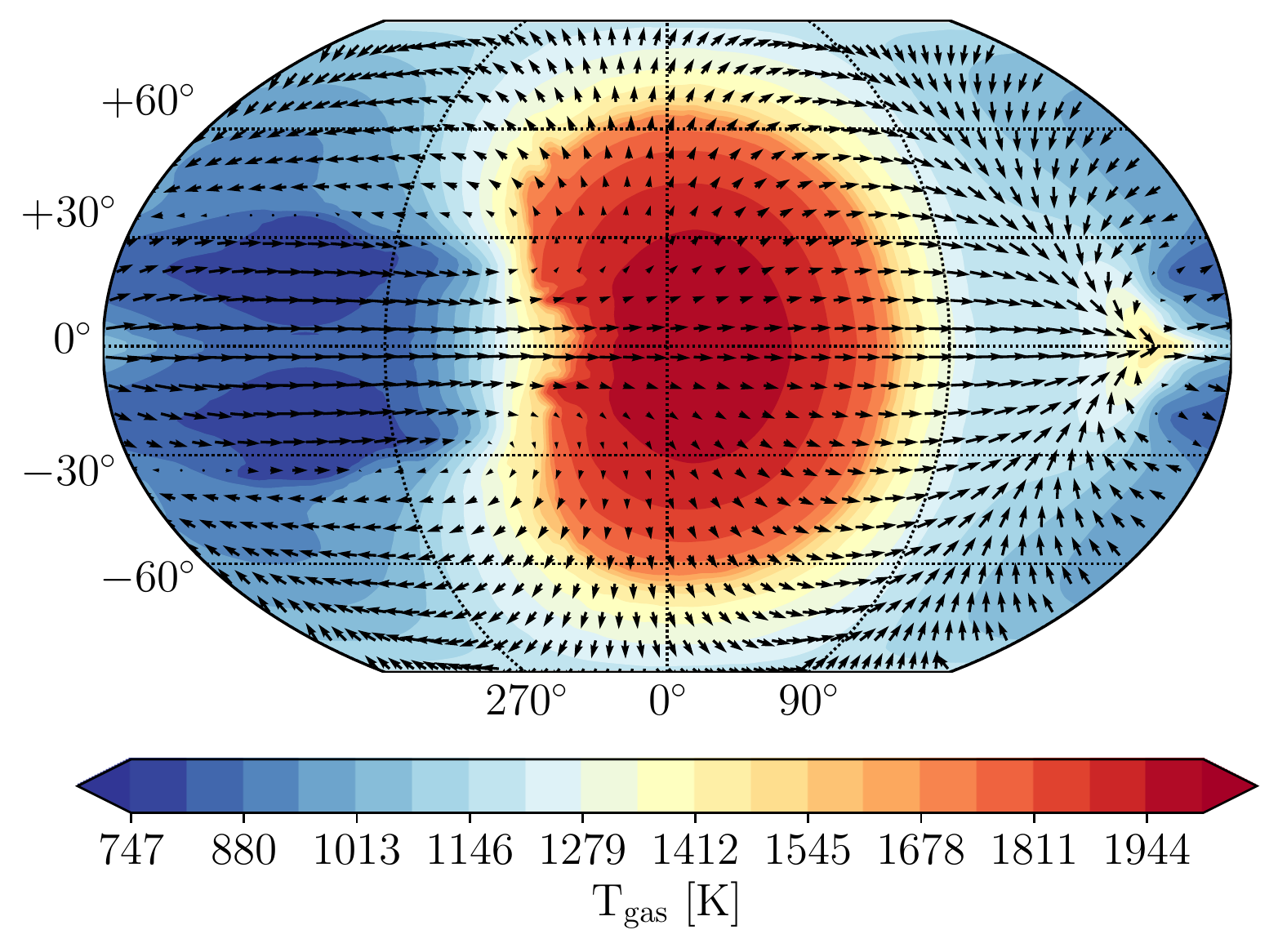}
   \includegraphics[width=0.49\textwidth]{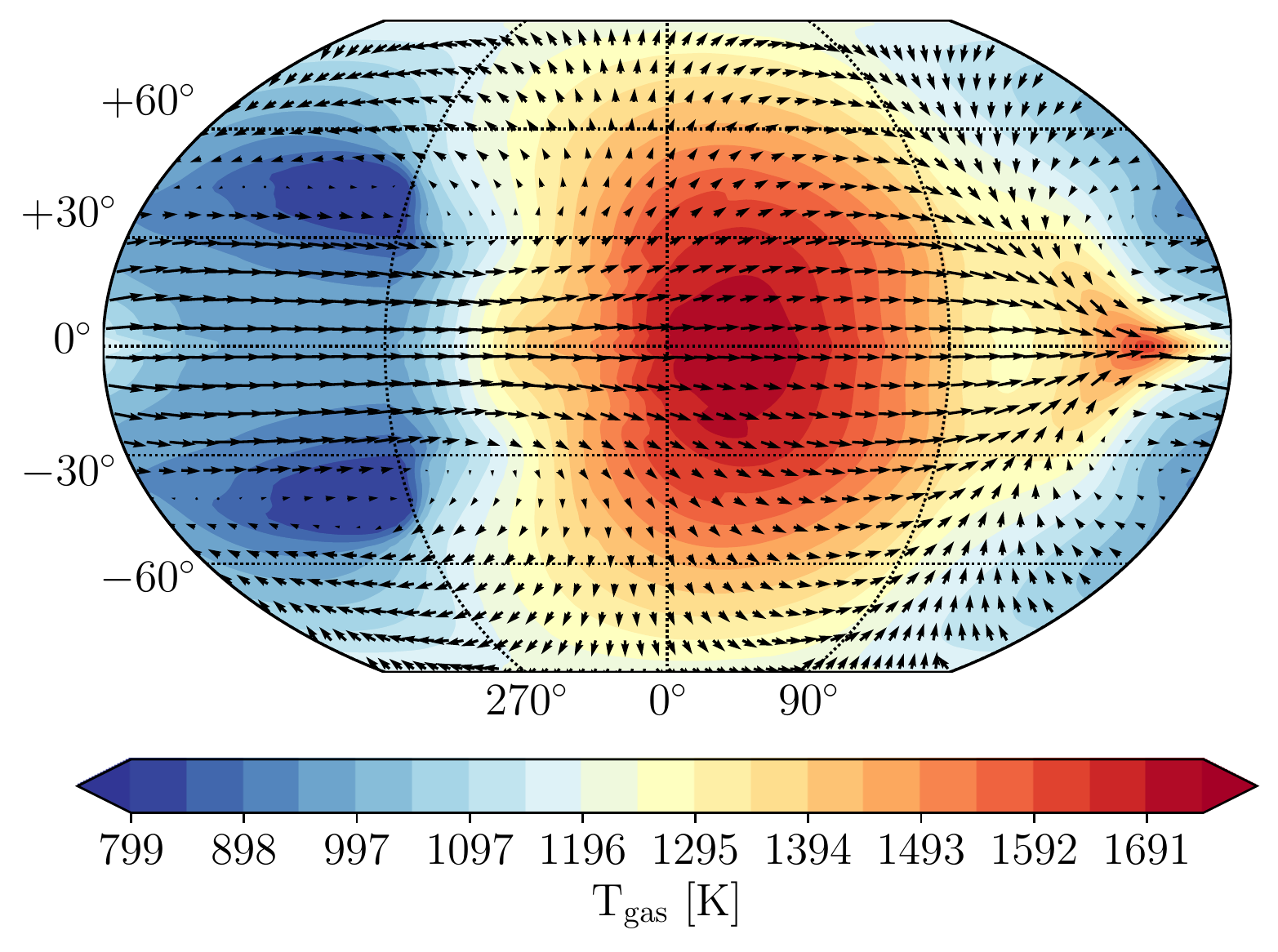}
   \includegraphics[width=0.49\textwidth]{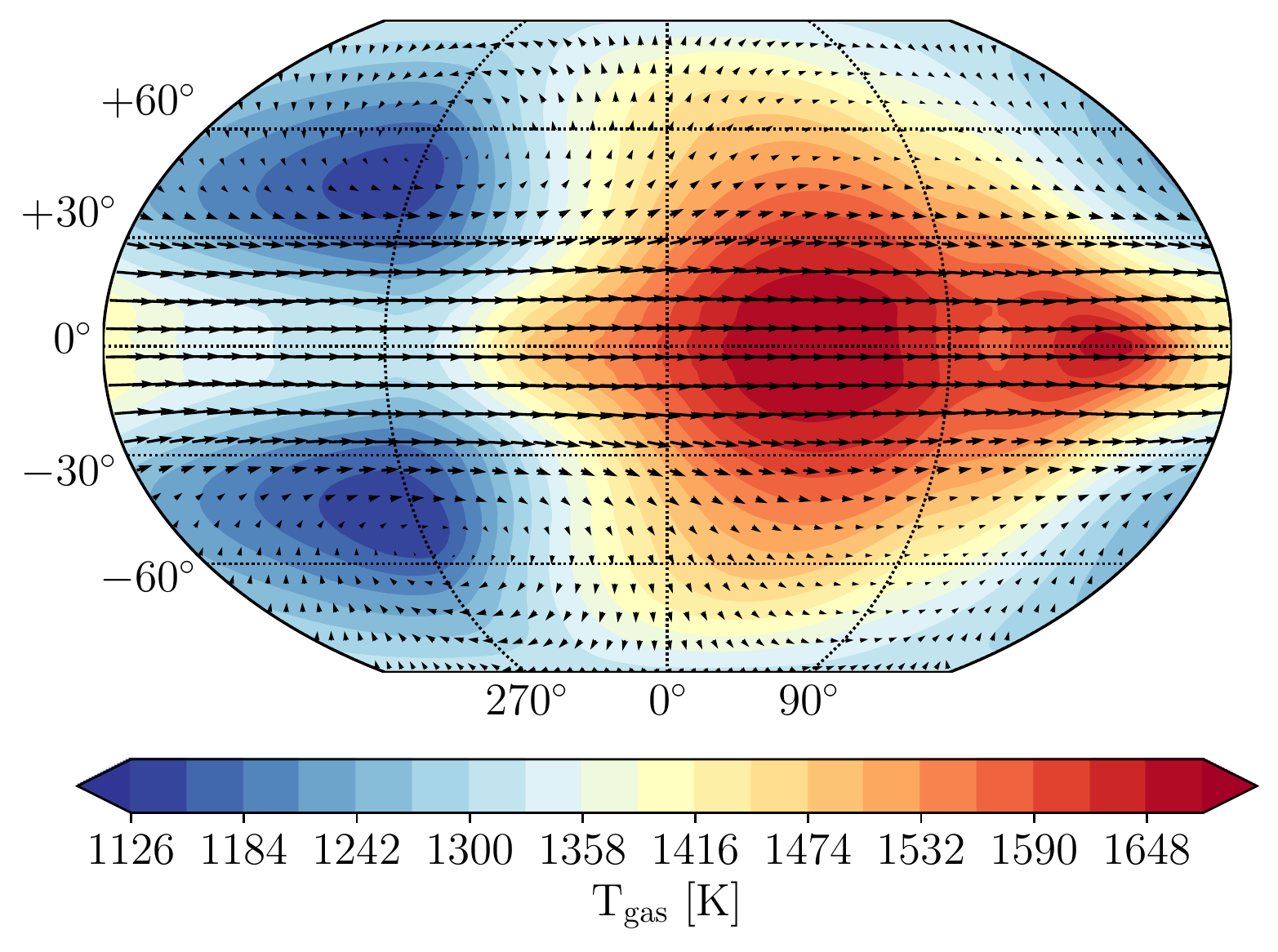}
   \includegraphics[width=0.49\textwidth]{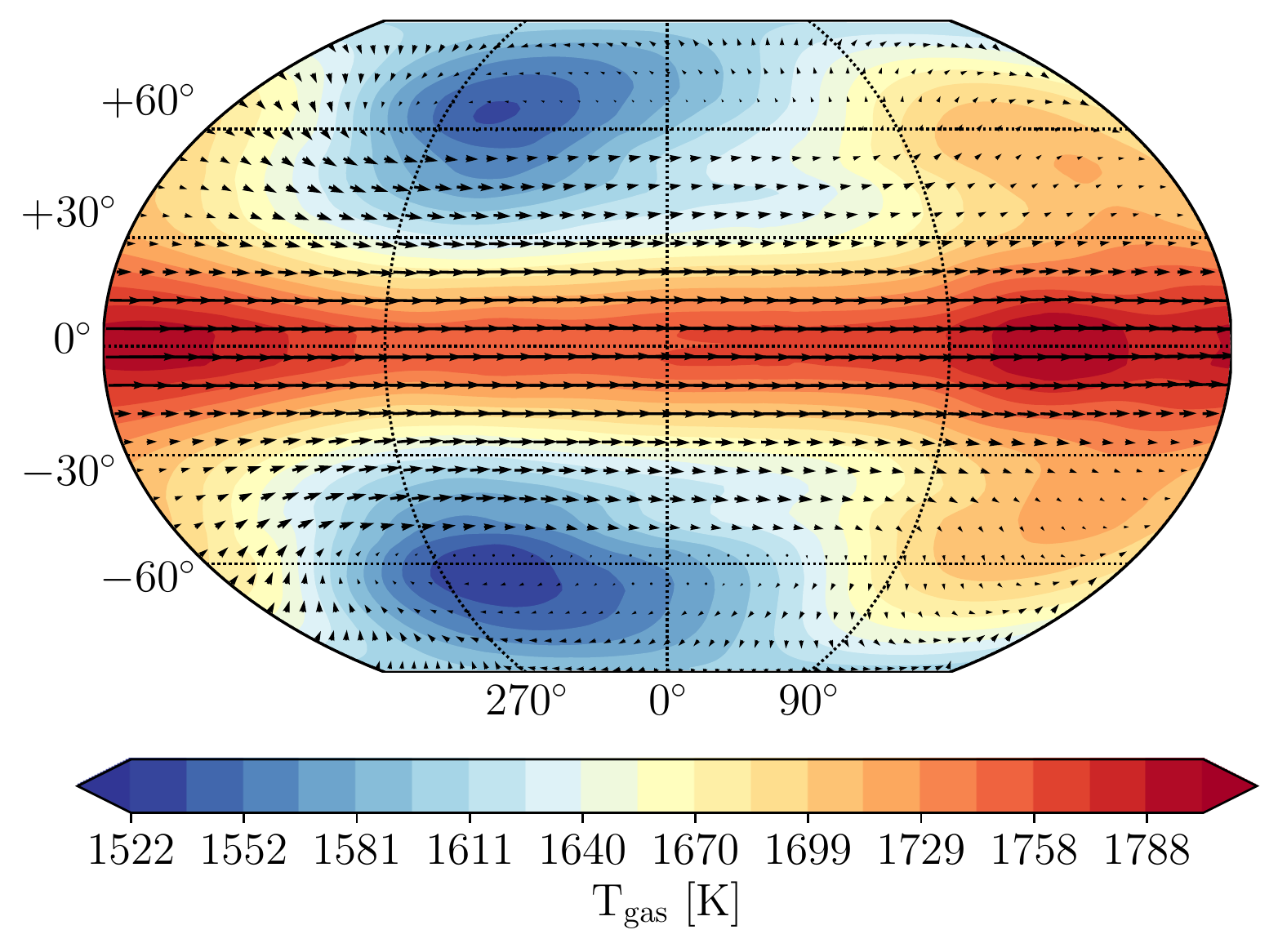}
   \caption{\textsc{Exo-FMS} output of a HD 209458b-like experiment using the correlated-k RT scheme.
   Top left: 1D equatorial T-p profiles (solid colour lines) and polar region (black dashed line).
   Top right: zonal mean zonal velocity.
   Middle left: lat-lon map of gas temperature at $\approx$1 mbar.
   Middle right: lat-lon map of gas temperature at $\approx$10 mbar.
   Bottom left: lat-lon map of gas temperature at $\approx$0.1 bar.
   Bottom right: lat-lon map of gas temperature at $\approx$1 bar.
   Velocity vectors are shown as back arrows.}
   \label{fig:HD209_spec}
\end{figure*}

We present results using \textsc{Exo-FMS} coupled to the real gas correlated-k model.
Due to the significant computational cost, the simulation is performed for 1600 days, with the final 100 day average taken as the presented results.
The simulation is initialised according to scheme outlined in Sect. \ref{sec:IC}.
For the input correlated-k opacities we use the pre-mixed tables with equilibrium condensation as outlined in Sect. \ref{sec:spec_RT}.
Figure \ref{fig:HD209_spec} shows the results of our coupled GCM and correlated-k scheme.

As seen in Fig. \ref{fig:HD209_spec} our T-p profiles show fluctuations in some dayside profiles at upper atmospheric regions.
This can attributed to the condensation of TiO and VO which happens near the T-p of the fluctuations leading to strong gradients in the VMR of gas phase TiO and VO across this T-p range.
An additional explanation is possibly related to numerical errors when interpolating the pre-mixed k-table T-p points to the local atmospheric T-p as discussed in \citet{Amundsen2017}.
This artefact can possibly be improved upon by increasing the k-table resolution near the temperatures important for TiO and VO condensation,
making a pre-mixed k-table without TiO and VO included (commonly used in the literature e.g. \citet{Amundsen2016, Kataria2016}), or using a random overlap scheme \citep{Amundsen2017}.

\subsection{HD 209458b-like comparisons}
\label{sec:comp}

\begin{figure*} 
   \centering
   \includegraphics[width=0.49\textwidth]{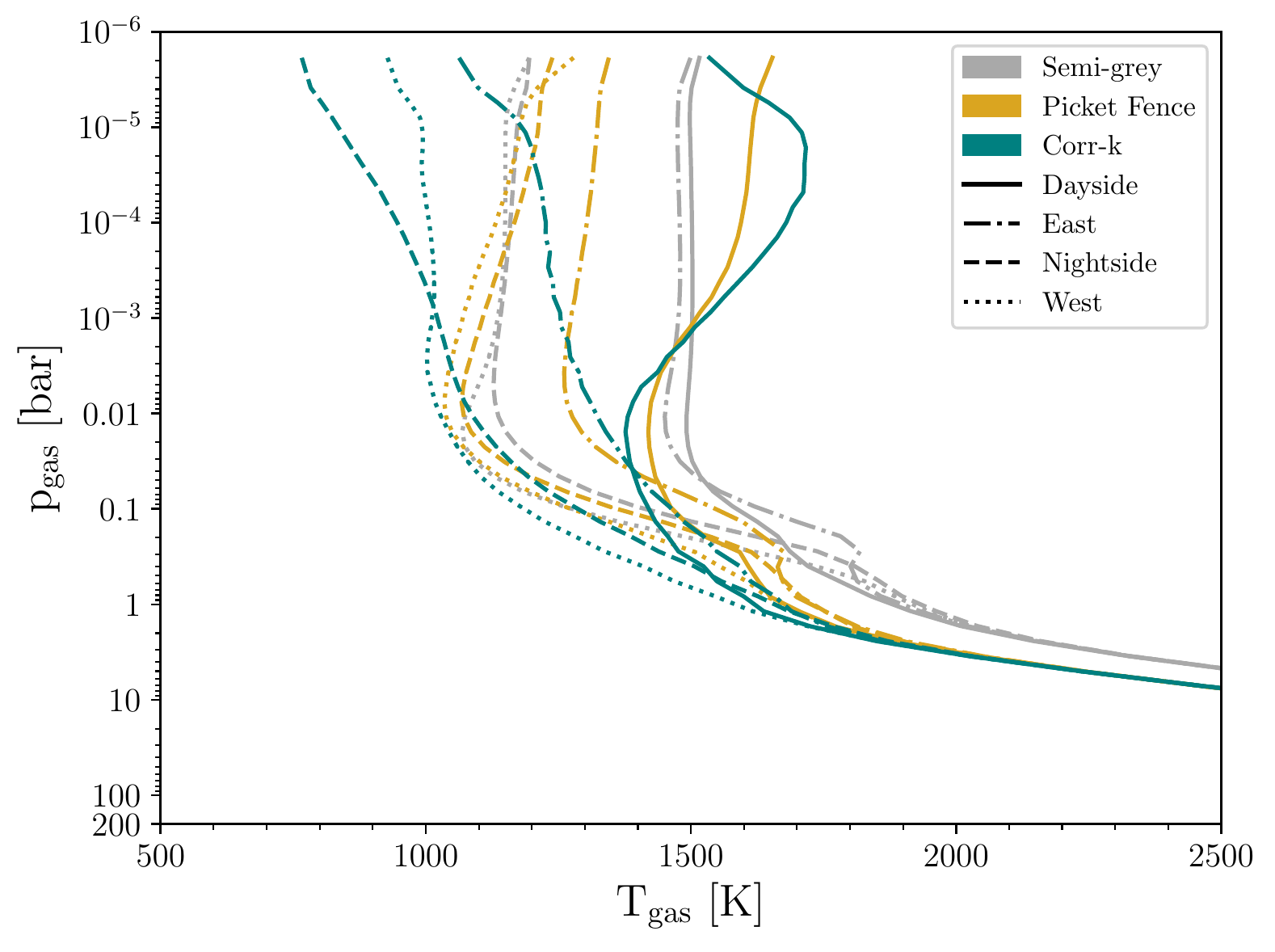}
   \caption{Mean T-p profile comparisons between the semi-grey, picket fence and corr-k HD 209458b-like simulations.
   Dayside, nightside, eastern terminator and western terminator mean T-p profiles are plotted in solid, dash-dot, dashed and dotted line styles respectively.}
   \label{fig:1D_means}
\end{figure*}

The correlated-k and picket fence scheme produce very similar zonal mean velocity structures and absolute values.
The temperature maps are also similar across the 10$^{-4}$-1 bar range.
Differences are mostly seen in the upper atmospheric temperatures at low pressure ($\lesssim$ 10$^{-4}$ bar), where the 30 correlated-k bands are able to more efficiently cool the atmosphere compared to the two band picket fence scheme.
The upper atmospheric temperature inversion and maximum temperature due to TiO and VO opacities are also well reproduced by the picket fence scheme.
Since the radiative-timescales are short at these low pressures, these differences are most likely a direct result of the RT scheme used, rather than any dynamical effect.

Due to the added computational burden when performing correlated-k models, our chosen end points between the correlated-k, picket fence/semi-grey runs were different (1600 and 3600 days respectively).
We have checked the daily output of the non-grey and semi-grey models at 1600 days, which shows a small difference (at most $\lesssim$ 5$\%$, within the natural variability of the model) in the T-p profiles, lat-lon maps and zonal mean velocities from the 100 day average at 3600 days.
This suggests that the extra 2000 days runtime in the non-grey and semi-grey models does not affect the outcome of the post-processing much, and the comparison to the corr-k output is still reasonable to perform.

The semi-grey model produces a different upper atmospheric temperature structure compared to the correlated-k model, with temperature patterns in the at pressures $<$ 10$^{-1}$ bar appearing notably different.
The central jet is also more extended in latitude compared to the picket fence and correlated-k models.

In Fig. \ref{fig:1D_means} we compare the average dayside, nightside and east and west terminator regions from each HD 209458b simulation.
It is clear that the correlated-k and picket fence model are similar, especially in the higher pressure regions.
Also evident is the different deep adiabatic gradient between the semi-grey model and the picket fence and corr-k models.
This is probably due to assuming a too strong exponent term scaling in the semi-grey optical depth parameters.
It is likely the pressure dependent semi-grey parameters could be adjusted to fit the adiabatic region produced in the corr-k model better, however, it is not clear how to set these parameters on a case by base basis without first comparing directly to the corr-k or picket fence model results.

In comparison to previous studies, we are able to reproduce qualitatively well the HD 209458b simulations performed in \citep{Showman2009}, with a similar range of temperature and pressures with depth and lat-lon pressure level maps.
Significant differences in the deep T-p structures are due to a lower internal temperature used in \citet{Showman2009} and different initial condition assumptions used in this study.
Our results are less compatible with the simulations performed \citet{Kataria2016} and \citet{Amundsen2016} due to our inclusion of TiO and VO opacities which greatly affect the dayside T-p profiles.
However, our nightside profiles show good agreement with the results of \citet{Kataria2016} and \citet{Amundsen2016}, suggesting our correlated-k scheme is producing comparable results to other well used models.

\section{Post-processing}
\label{sec:pp}

\begin{figure*} 
   \centering
   \includegraphics[width=0.49\textwidth]{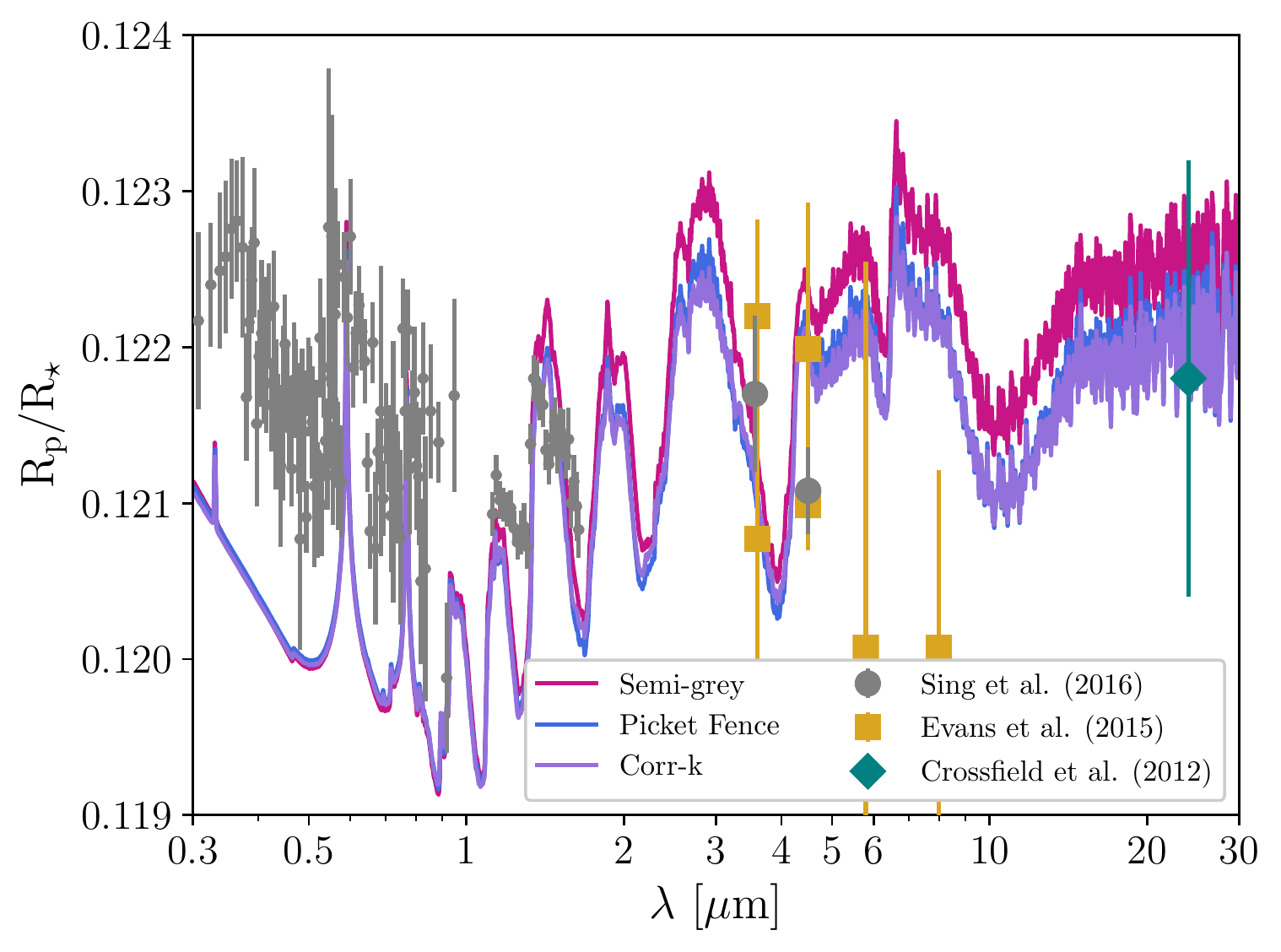}
   \includegraphics[width=0.49\textwidth]{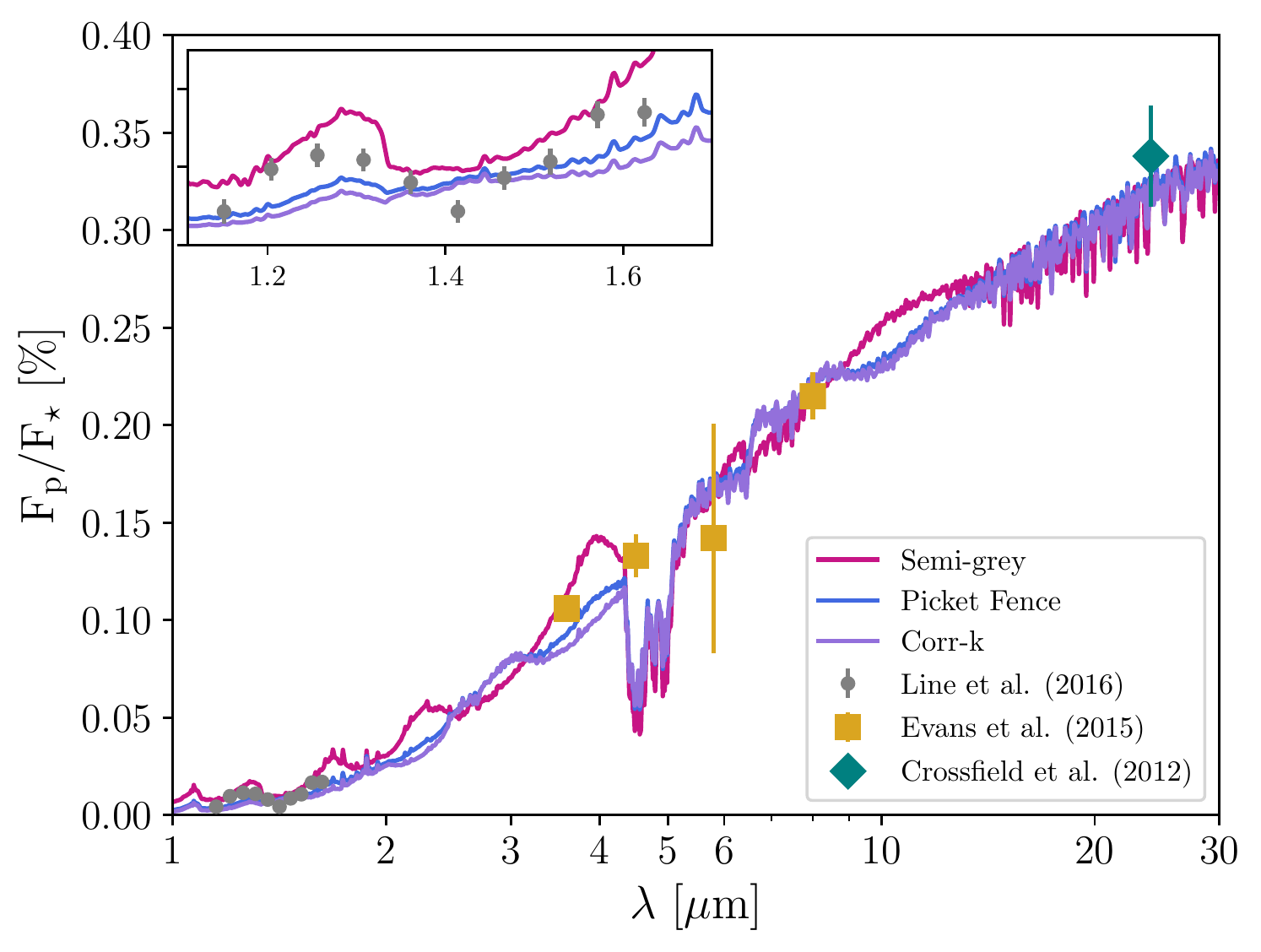}
   \caption{Transmission and dayside emission spectra post-processing of the HD 209458b simulations.
   For the transmission spectra, the \citet{Sing2016} (grey points), \citet{Evans2015} (gold squares) and \citet{Crossfield2012} (teal diamond) observations are over-plotted for comparison.
   For the dayside emission spectra, the \citet{Line2016} (grey points), \citet{Evans2015} (gold squares) and \citet{Crossfield2012} (teal diamond) observations are also plotted.}
   \label{fig:pp}
\end{figure*}

\begin{figure*} 
   \centering
   \includegraphics[width=0.49\textwidth]{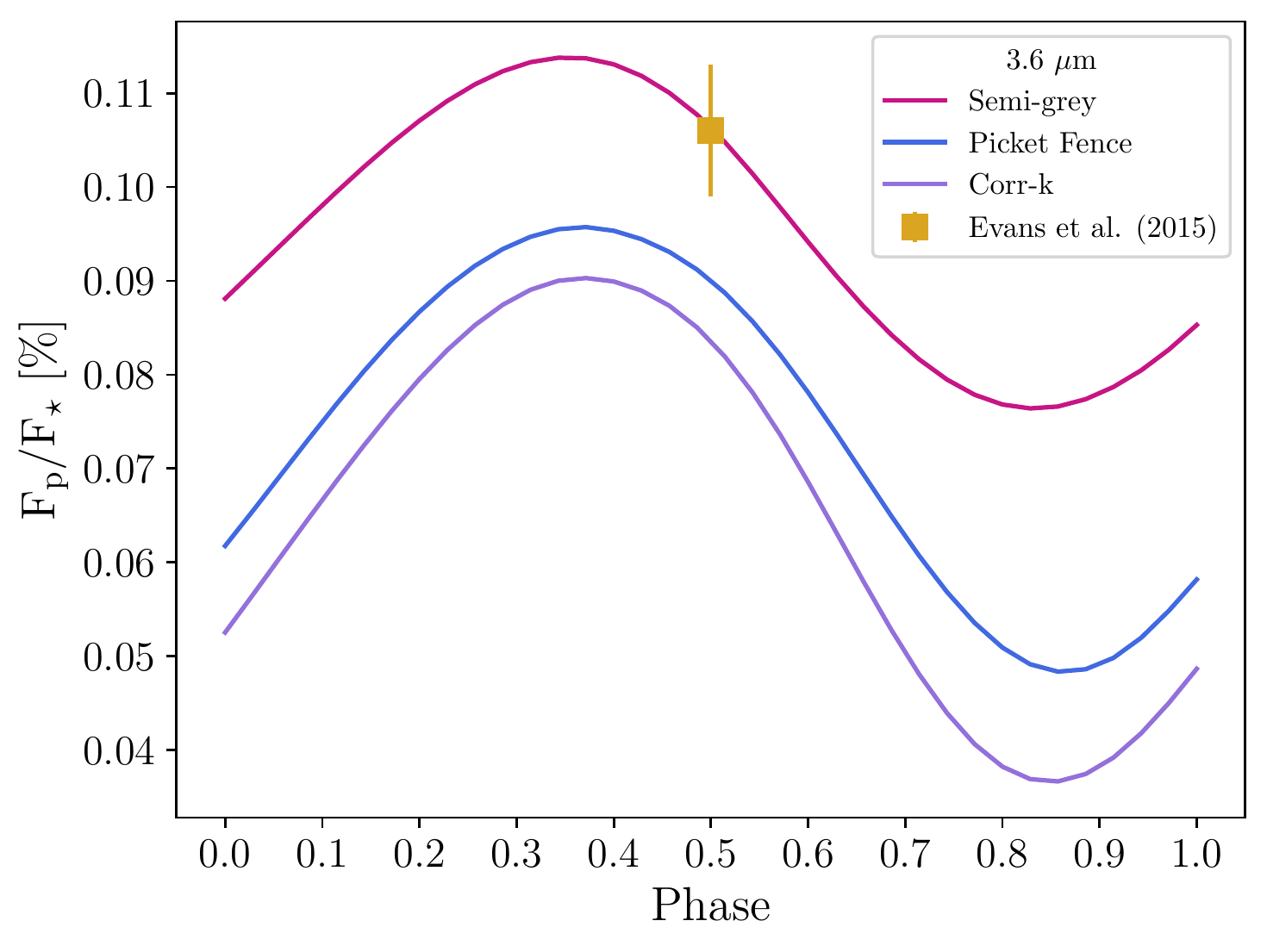}
   \includegraphics[width=0.49\textwidth]{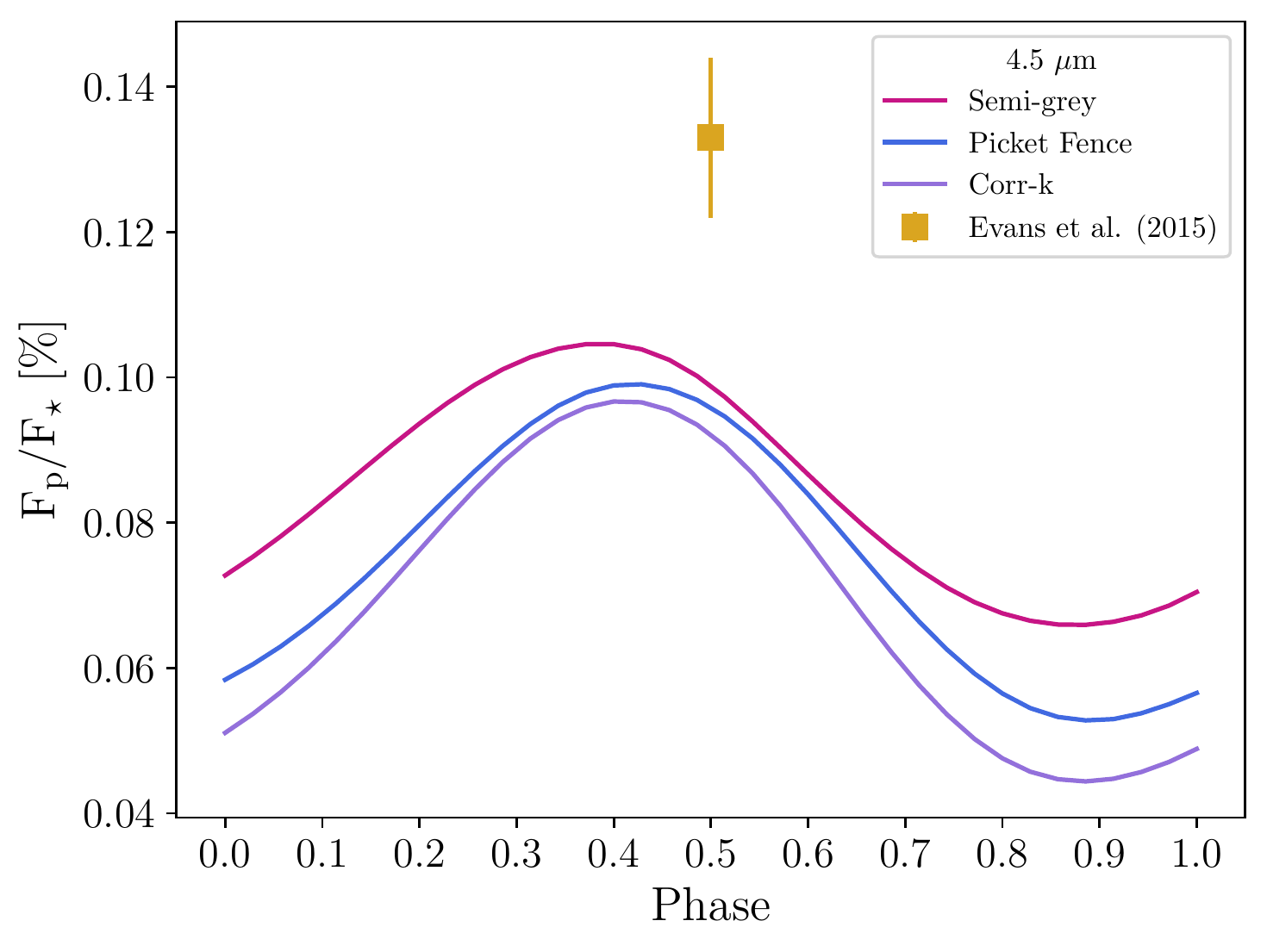}
   \includegraphics[width=0.49\textwidth]{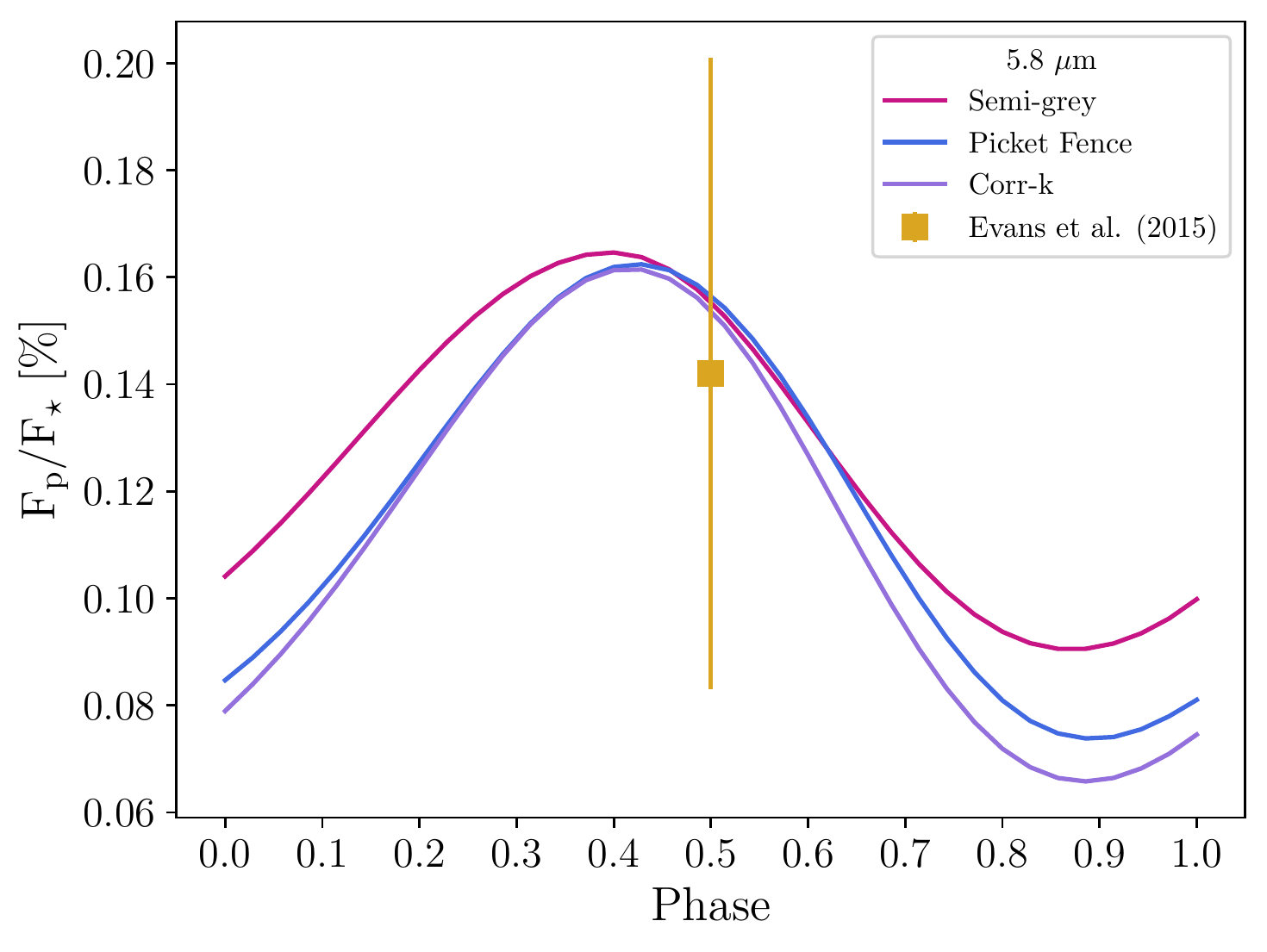}
   \includegraphics[width=0.49\textwidth]{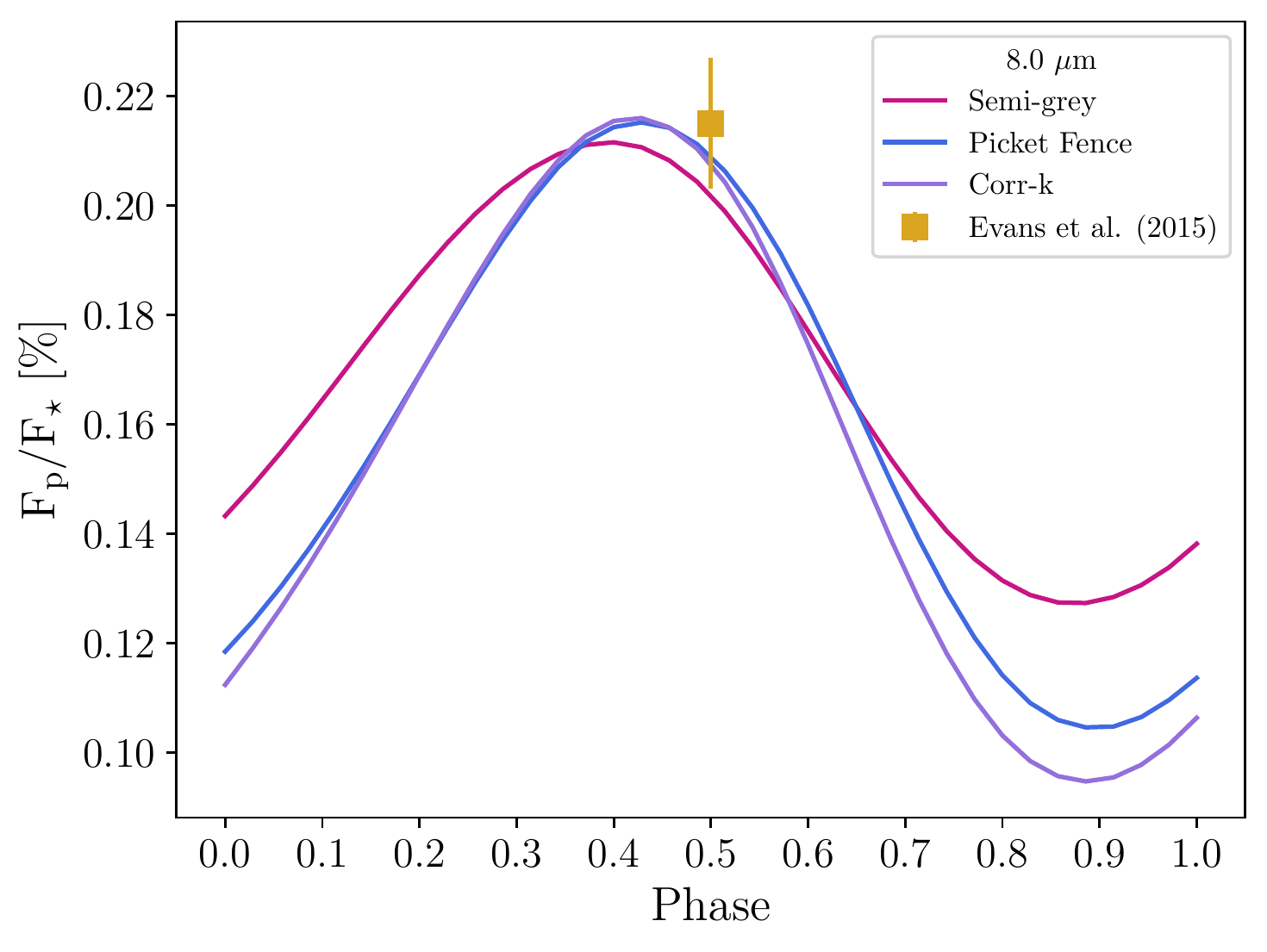}
   \includegraphics[width=0.49\textwidth]{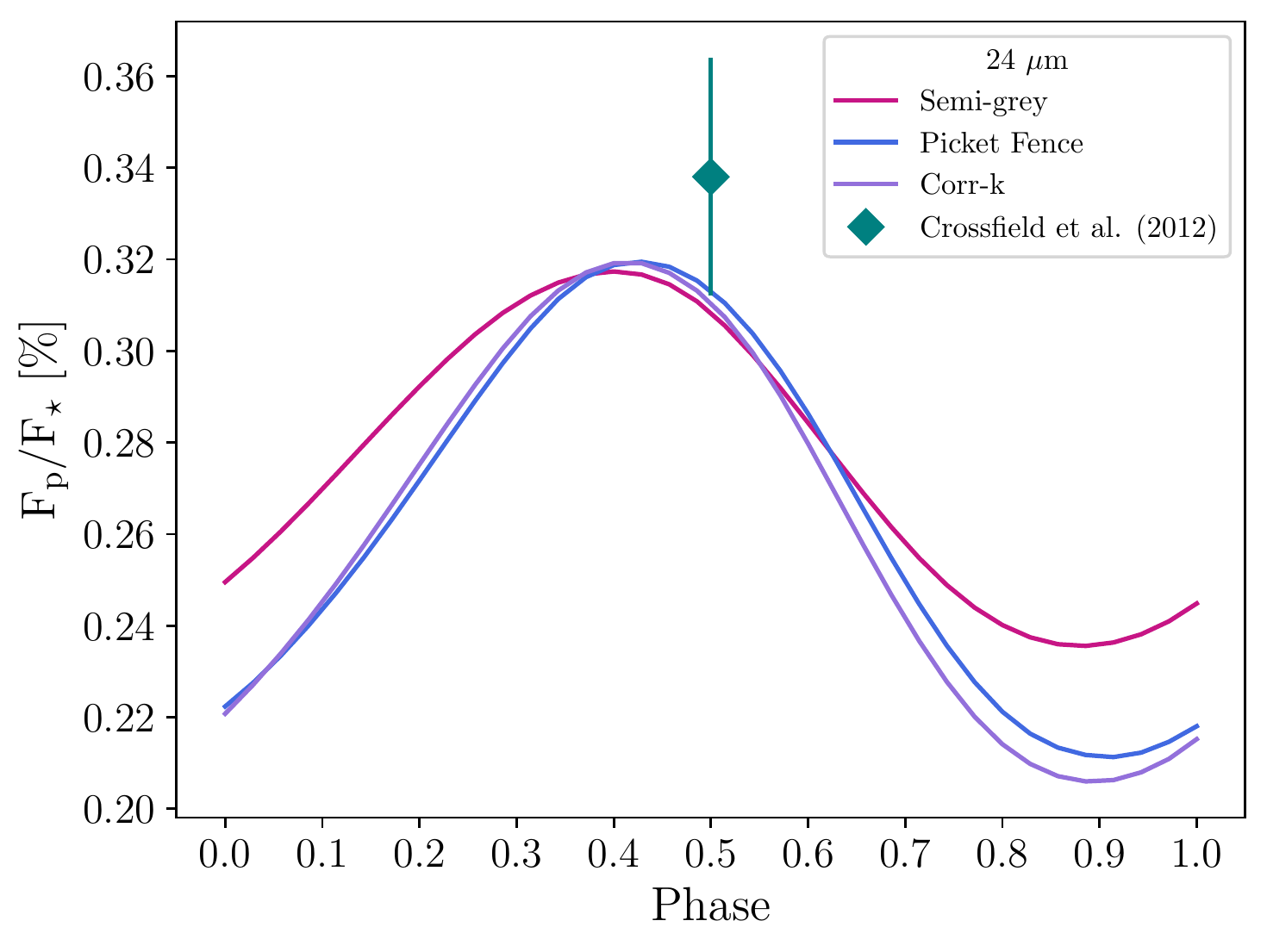}
   \caption{Phase curves for Spitzer IRAC and MIPS bands post-processing of the HD 209458b simulations.
  The dayside emission values from \citet{Evans2015} and \citet{Crossfield2012} are plotted at 0.5 phase.}
   \label{fig:ph}
\end{figure*}

In this section we post-process our HD209458b-like simulations and compare the results to available observational data.
We use the CMCRT code of \citet{Lee2017} in correlated-k mode \citep{Lee2019}.
For CMCRT input opacities we use \textsc{nemesis} \citep{Irwin2008} formatted k-tables available from the \textsc{ExoMolOP} database \citep{Chubb2021}.
Table \ref{tab:line-lists} shows the opacity sources and references used for the post-processing.

Figure \ref{fig:pp} presents the transmission and dayside emission spectra.
For the transmission spectra, the correlated-k model was scaled to the $\approx$1.4 $\mu$m H$_{2}$O feature and the picket fence and grey gas results scaled to the Rayleigh scattering slope of the real gas model.
From the transmission spectra figure it is clear that the picket fence and corr-k model produce highly similar transmission spectra.
Our semi-grey model produces larger molecular features, probably due to the increased scale height due it's generally higher atmospheric temperatures.
In emission, again the picket fence and corr-k model agree well across the whole wavelength range, with the most difference occurring around the 4$\mu$m range.

Figure \ref{fig:ph} presents a comparison of the Spitzer bandpass phase curves predictions between the models.
From these results it is clear the picket fence scheme reproduces well the correlated-k model results, with good agreement for the maximum flux shift and day-night contrast.
However, the picket fence scheme has more difficulty reproducing the nightside fluxes of the corr-k model, where the cooling of the upper atmosphere is more important in controlling the T-p structure.

\section{Discussion}
\label{sec:discussion}

\begin{figure*} 
   \centering
   \includegraphics[width=0.49\textwidth]{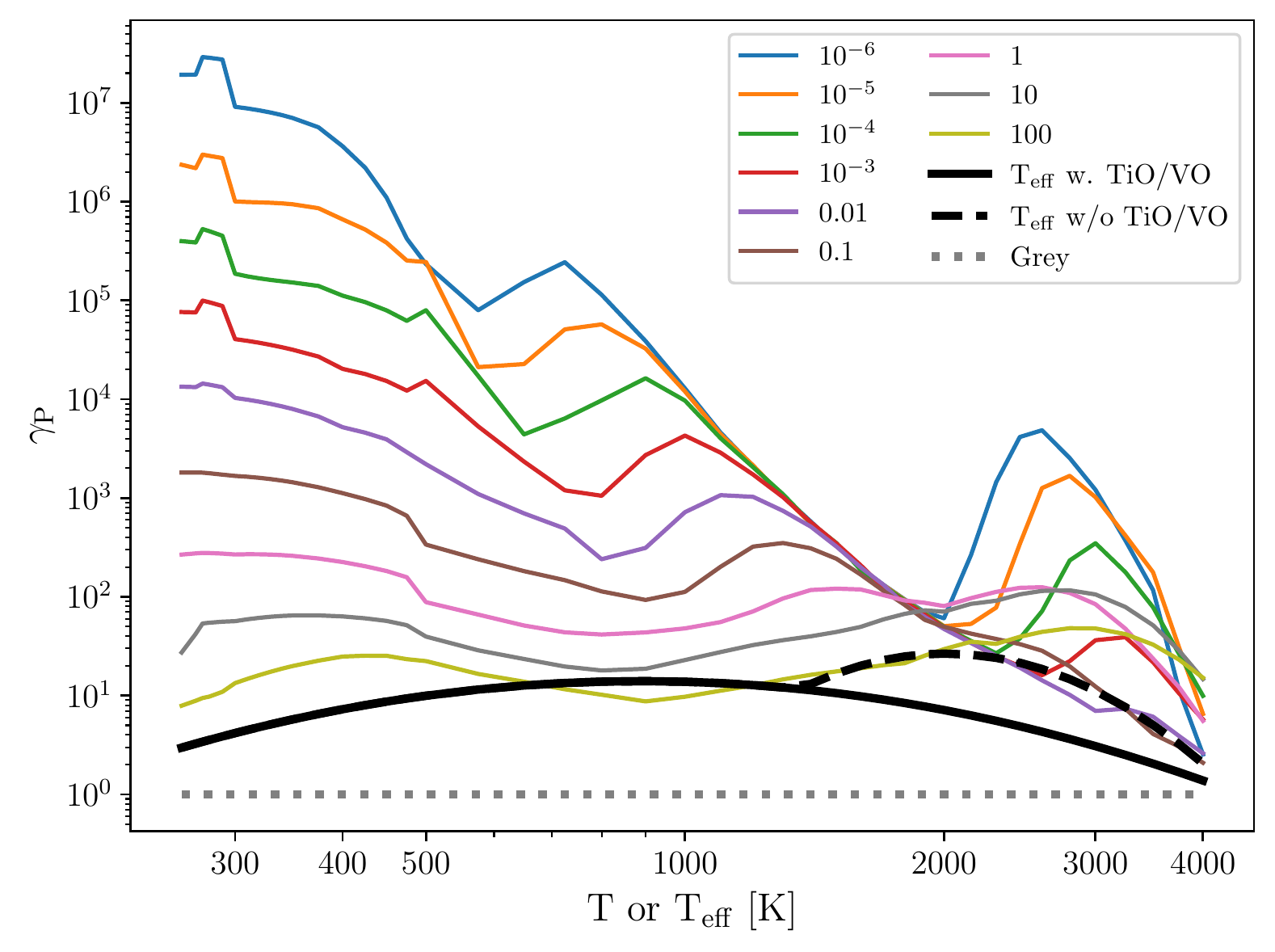}
   \caption{$\gamma_{P}$ = $\kappa_{P}$/$\kappa_{R}$ calculated from the solar metallicity \citet{Freedman2014} tables.
   Coloured lines denote the gas pressure in bar.
   The black lines show the \citet{Parmentier2015} expression as a function of T$_{\rm eff}$ (solid: with TiO/VO, dashed: w/o TiO/VO).
   The dotted grey line shows the grey region where $\kappa_{P}$ = $\kappa_{R}$.}
   \label{fig:non-grey}
\end{figure*}

Our picket fence scheme following \citet{Parmentier2014} and \citet{Parmentier2015} shows promise in improving the realism of the RT calculations with simple modifications to current semi-grey RT schemes.
Our initial modelling was able to produce good agreement with the dayside temperature profiles of real gas model.
We caution a more quantitive agreement will require a more careful choice with respect to the best parameterisations with this scheme.
Furthermore, our test was not a fair one-to-one comparison due to several factors:
\begin{itemize}
\item Different opacity sources between the correlated-k scheme and Rosseland mean \citet{Freedman2014} studies.
\item Different gas phase abundances through the choice of solar elemental ratio reference, different CE solver and equilibrium condensation data used between \citet{Freedman2014} and \citet{Woitke2018}.
\end{itemize}
As seen in our GCM results and also discussed in 1D models of \citet{Parmentier2015}, the most difference between the picket fence and real gas models occurs in the upper atmosphere from p$_{\rm gas}$ $\lesssim$ 10$^{-3}$ bar.
This is due the picket fence scheme not actively capturing efficient cooling that occurs in these less dense regions, as extensively proven in \citet{Parmentier2014,Parmentier2015}.

However, given the simplicity of implementing a picket fence scheme, this offers a useful `middle-ground' in physical realism between the semi-grey and real gas RT schemes.
This scheme has potential to be further generalised with the inclusion of shortwave and longwave scattering components \citep[e.g.][]{Pierrehumbert2010, Mohandas2018}, enabling the simulation of radiative feedback due to cloud formation.
Including radiative feedback from non-equilibrium chemistry may be more difficult to include in the current picket fence framework, since the Rosseland mean tables from \citet{Freedman2014} use chemical equilibrium results.
A more generalised scheme would be required to take varying volume mixing ratios into account.
A potential hybrid approach may be to evolve simulations for an extended `spin-up' period using the picket fence scheme, switching to the correlated-k scheme for the latter half of the simulation when adding additional feedback mechanisms.

Our current simulations are not representative of a fully converged simulation due to long radiative-timescales and mixing timescales (10000+ days) in the lower atmosphere, examined by recent studies \citep[e.g.][]{Mayne2017,Wang2020}.
We note that a long integration time on the scale of \citet{Wang2020} is unfeasible for scope of this study.
Despite this limitation of the current study, the proposed picket fence scheme offers an efficient and more realistic RT solution than semi-grey schemes for longer integration investigations to study deep layer radiative and momentum transport in the atmosphere.

The `greyness' of the wavelength dependent opacity structure can be inferred by the ratio of the Planck mean to Rosseland mean \citep[e.g.][]{Guillot2010}
\begin{equation}
\gamma_{P} = \frac{\kappa_{P}}{\kappa_{R}},
\end{equation}
where if $\gamma_{P}$ = 1, the atmospheric opacity is grey with $\gamma_{P}$ $\gg$ 1 characteristic of a non-grey atmosphere.
In Fig. \ref{fig:non-grey} we show $\gamma_{P}$ as given by the Planck and Rosseland mean, solar metallicity opacity tables presented in \citet{Freedman2014}.
This shows that generally the non-greyness of an atmosphere increases with decreasing temperature and pressure, typical environments important for the atmospheric cooling efficiency.
In addition, we show the $\gamma_{P}$ expression from \citet{Parmentier2015} with and without TiO and VO opacity included.
This shows that the relationships between the parameters in \citet{Parmentier2015} is vital to accurate determination of the visible band and picket fence opacity ratios for the column, since putting directly the local $\gamma_{P}$ from the \citet{Freedman2014} tables into the framework would probably not result in fitting the correlated-k T-p profiles well.

\section{Summary \& Conclusions}
\label{sec:conclusions}

In this study we successfully benchmarked the \textsc{Exo-FMS} GCM model to contemporary studies of hot gas giant model atmospheres.
Using the semi-grey scheme, we were able to reproduce qualitatively and quantitatively the results of the \citet{Heng2011b} study, but only compared qualitatively with the \citet{Rauscher2012} results.
We coupled a real gas, correlated-k scheme to \textsc{Exo-FMS}, and were able to match well with the general behaviour of \citet{Showman2009}, although with differences in the detailed results.

We propose a general initial condition scheme using the analytical radiative-equilibrium profiles from \citet{Parmentier2014,Parmentier2015} at the sub-stellar point.
This offers a simple way to practically implement recent recommendations by \citet{Sainsbury-Martinez2019} and internal temperature considerations in \citet{Thorngren2019}.

We implemented a non-grey picket fence RT scheme \citep{Parmentier2014,Parmentier2015} in the GCM and compared the results of the semi-grey, picket fence and real gas RT schemes on a HD 209458b-like simulation set-up.
The picket fence scheme was able to reproduce well the global and local T-p profiles of the real gas model at pressures most responsible to the observable characteristics of the atmosphere.
The picket fence model produces highly comparable transmission and emission spectra to the corr-k model when post-processed.
The phase curve properties of the picket fence model are also very comparable to the corr-k results, with the most differences occurring on the nightside regions of the planet.

Overall, we suggest that the picket fence approach offers a potentially efficient way to significantly improve the realism of the radiative-transfer characteristics in hot Jupiter GCMs, avoiding use of computationally expensive real gas calculations.
We suggest that the implementation of such schemes will be highly useful for future GCM modelling that require longer integration times or experiments with parameter examinations that wish to retain a more realistic RT solution, or are interested in producing computationally cheaper GCM results while retaining the transmission, emission and phase curve characteristics of the atmosphere given by the more expensive RT schemes.

\section*{Data and code availability}
1D column versions that emulate the FMS GCM of the semi-grey, picket fence schemes and correlated-k approach are available from the lead author's GitHub: \url{https://github.com/ELeeAstro}
These include our initial condition module and dry convective adjustment scheme.
GCM model output in NetCDF format is available from Zendoo, DOI: 10.5281/zenodo.5011921 .
All other data and code is available from the authors on a collaborative basis.

\section*{Acknowledgements}
We thank N. Lewis and M. Line for advice on RT schemes.
E.K.H. Lee is supported by the SNSF Ambizione Fellowship grant (\#193448).
This work was supported by European Research Council Advanced Grant \textsc{exocondense} (\#740963).
Plots were produced using the community open-source Python packages Matplotlib \citep{Hunter2007}, SciPy \citep{Jones2001}, and AstroPy \citep{Astropy2018}.
The HPC support staff at AOPP, University of Oxford and University of Bern are highly acknowledged..




\bibliographystyle{mnras}
\bibliography{bib2} 



\clearpage

\appendix

\section{RT scheme validation}
\label{app:RT_validation}

\begin{figure*} 
   \centering
   \includegraphics[width=0.49\textwidth]{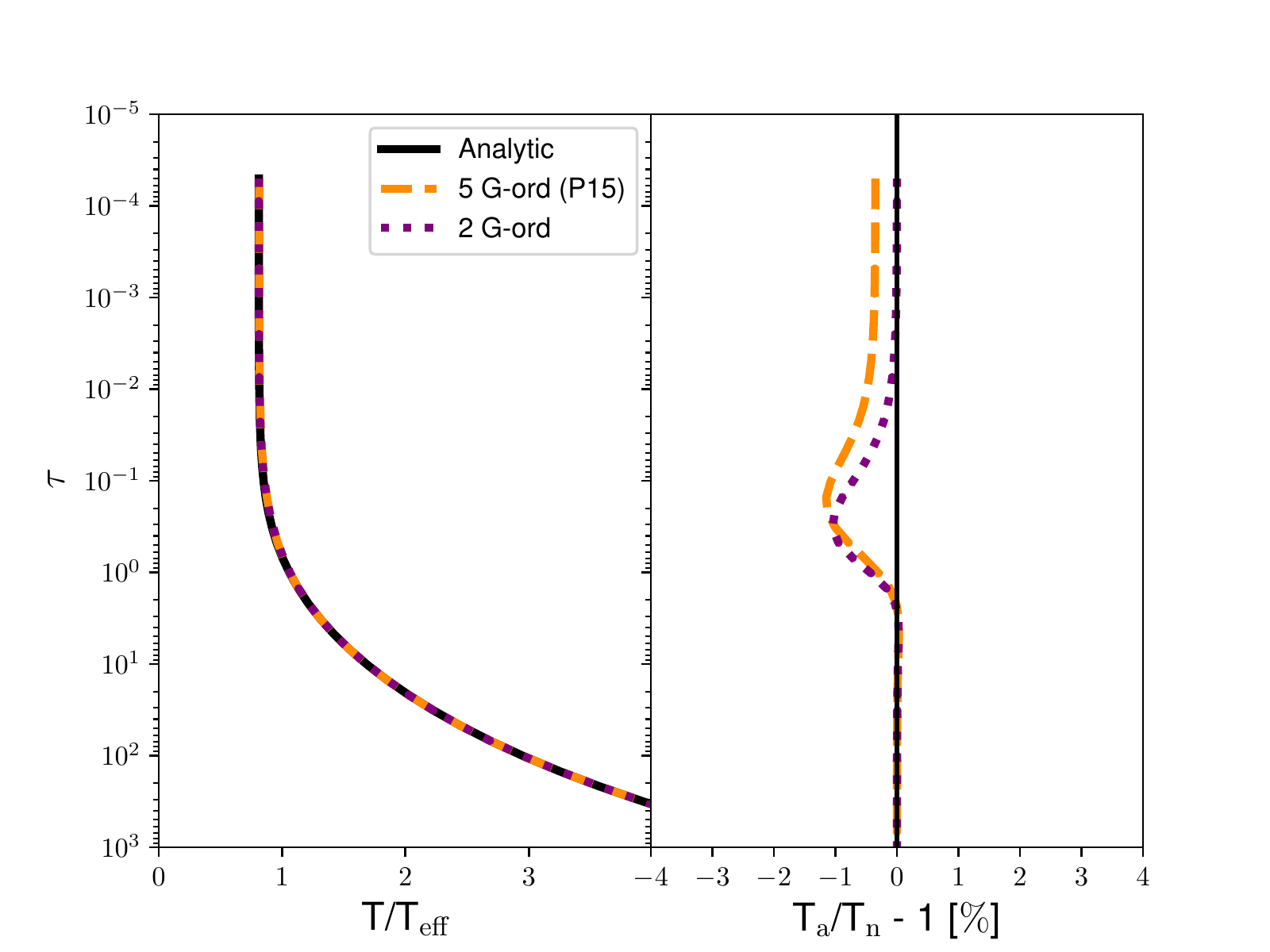}
   \includegraphics[width=0.49\textwidth]{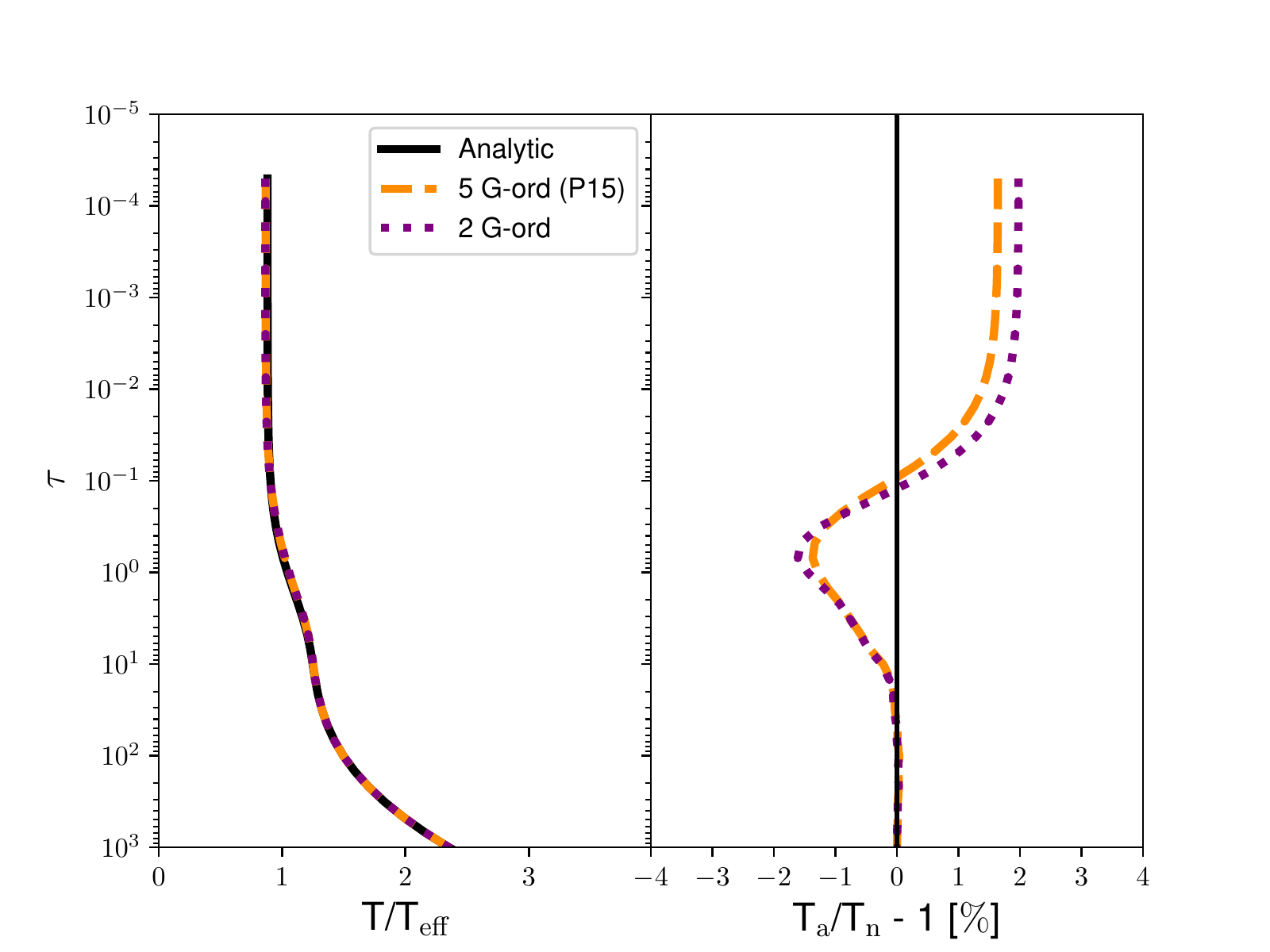}
   \caption{Validation of the semi-grey RT set-up following \citet{Parmentier2015}.
   Left: The analytical self-luminous solution of \citet{Chandrasekhar1960} (T$_{\rm a}$) is compared to the numerical solution (T$_{\rm n}$).
   Right: Semi-grey analytical T-p solution from of \citet{Guillot2010} (T$_{\rm a}$ - T$_{\rm irr}$ = 1288 K, T$_{\rm int}$ = 500 K, $\mu_{\star}$ = 1/$\sqrt{3}$, $\gamma_{\rm v}$ = 0.25) compared to the numerical solution (T$_{\rm n}$).
   In each case, we also compare to the 5 Gaussian quadrature points solution used in \citet{Parmentier2015} (P15), showing our 2 points give similar $\%$ error to the 5 point case.}
   \label{fig:SG_compare}
\end{figure*}

\begin{figure*} 
   \centering
   \includegraphics[width=0.49\textwidth]{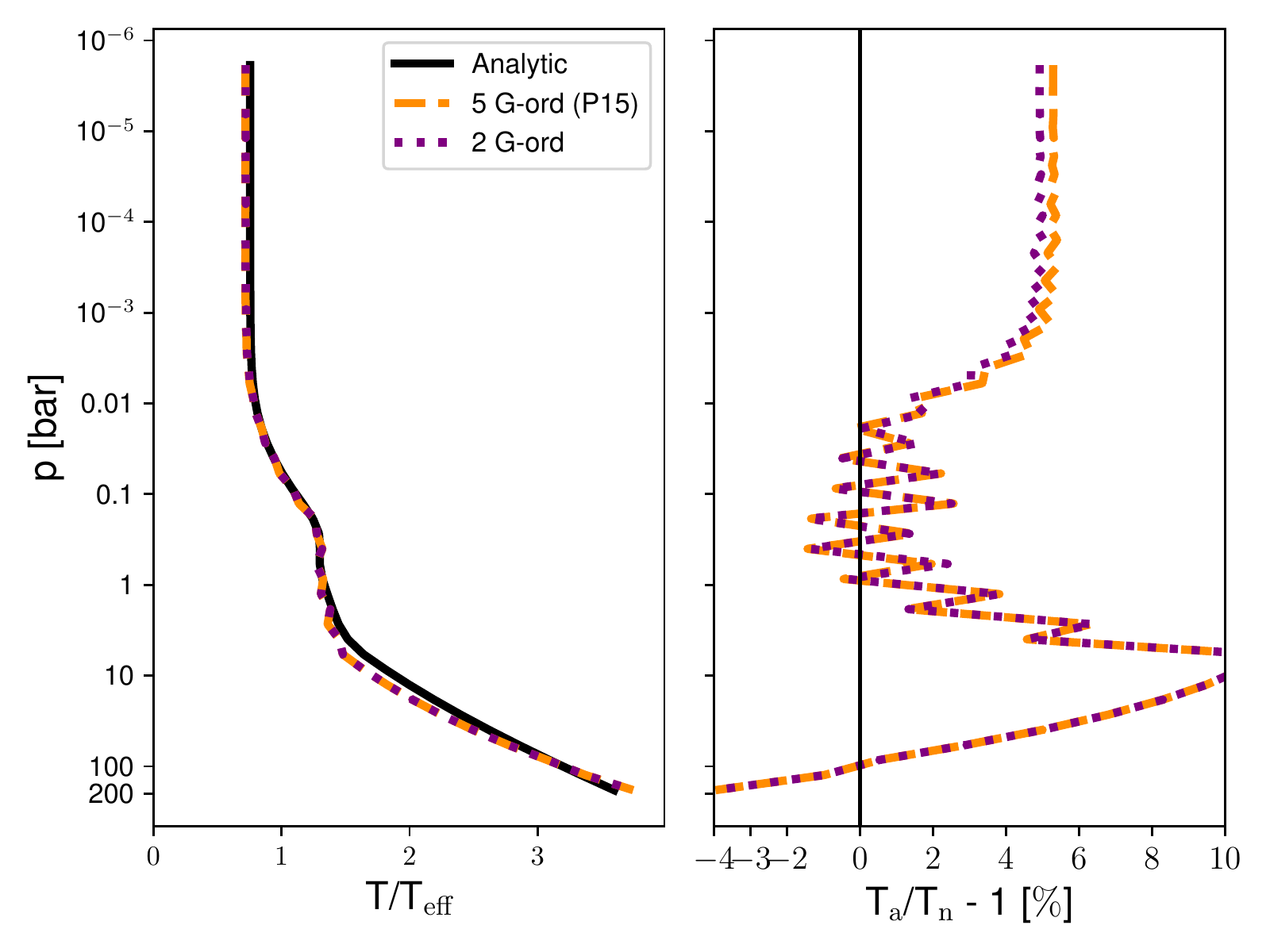}
   \includegraphics[width=0.49\textwidth]{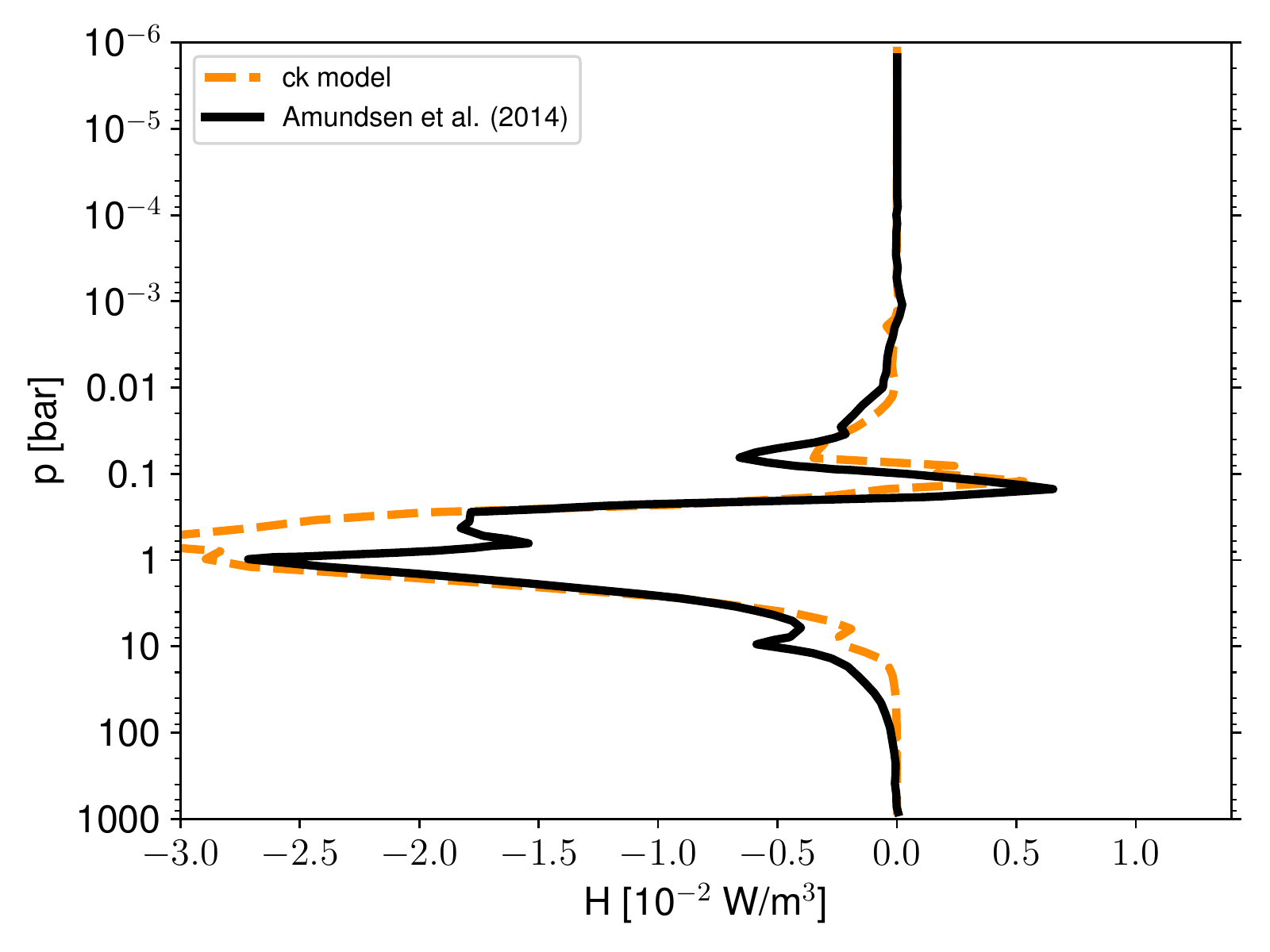}
   \caption{Validation of the semi-grey and corr-k RT set-up.
   Left: The analytical T-p  (T$_{\rm a}$ - T$_{\rm irr}$ = 1288 K, T$_{\rm int}$ = 500 K, $\mu_{\star}$ = 1/$\sqrt{3}$), following the analytical profile of \citet{Parmentier2015} and the equivalent numerical set-up.
   Right: The correlated-k model instantaneous heating rates, following the Test 2 T-p profile in \citet{Amundsen2014} with the scheme used in the GCM.}
   \label{fig:PF_compare}
\end{figure*}

In this appendix we validate the radiative-transfer scheme used in this study.
Figure \ref{fig:SG_compare} shows a comparison between the \citet{Chandrasekhar1960} analytic solution for a self-luminous atmosphere and a repeat of the \citet{Parmentier2015} for a \citet{Guillot2010} analytical semi-grey profile.
This shows our scheme produces excellent agreement to within 2 $\%$ of the analytical solutions.
Figure \ref{fig:PF_compare} shows a comparison with the picket fence scheme, showing good agreement with the analytical solution, to within 5$\%$ in the radiative region and 10$\%$ in the deeper convective region.
We also show heating rates from Test 2 in \citet{Amundsen2014}, showing excellent agreement for the correlated-k scheme to contemporary models.
Differences can most probably be attributed to the use of different line-list opacities between \citet{Amundsen2014} and this study, and differences between the analytic \citet{Burrows1999} CE abundances used in \citet{Amundsen2014} and the numerical CE solver \citet{Woitke2018} used in this study.

\section{GCM simulation parameters}
\label{app:GCM_parameters}

\begin{table*}
\centering
\caption{Adopted GCM simulation parameters for the HD 209458b-like semi-grey (Sect. \ref{sec:semi-grey2}), non-grey (Sect. \ref{sec:non-grey}) and correlated-k (Sect. \ref{sec:spectral}) simulations.
Bulk parameters are derived from the values found in the exoplanet.eu database.}
\begin{tabular}{c c c c c l}  \hline \hline
 Symbol & semi-grey & non-grey & corr-k & Unit & Description \\ \hline
 T$_{\rm irr}$ & 2091 & 2091  & spectral & K & Irradiation temperature \\
 A$_{\rm B}$ & 0 & $\star$ & 0 & - & Bond albedo \citep[$\star$][]{Parmentier2015} \\
 T$_{\rm int}$ & 571 & 571 &  571 &  K & Internal temperature \citep{Thorngren2019} \\
 P$_{\rm 0}$ & 220 & 220 & 220 & bar & Reference surface pressure \\
 $\kappa_{V}$ & 6.13$\cdot$10$^{-4}$  & $\star$ 3 bands & 30 bands & m$^{2}$g$^{-1}$ &  Visible band opacity \citep[$\star$][]{Parmentier2015} \\
 $\kappa_{IR}$ & 1$\cdot$10$^{-3}$  &  $\star$ 2 bands & 30 bands &  m$^{2}$g$^{-1}$ &  Infrared band opacity \citep[$\star$][]{Parmentier2015} \\
 n$_{\rm L}$ & 2 & - & - & - & IR power-law index \\
 f$_{l}$ & 0.0005 & - & - & - & IR linear component fraction \citep{Heng2011b} \\
 c$_{\rm p}$ & 1.3$\cdot$10$^{4}$  & 1.3$\cdot$10$^{4}$ & 1.3$\cdot$10$^{4}$ & J K$^{-1}$ kg$^{-1}$ & Specific heat capacity \\
 R & 3556.8 & 3556.8  & 3556.8 & J K$^{-1}$ kg$^{-1}$  & Ideal gas constant \\
 $\kappa$ & 0.2736 & 0.2736 & 0.2736 & J K$^{-1}$ kg$^{-1}$  & Adiabatic coefficient \\
 g$_{\rm HJ}$ & 8.98 & 8.98 & 8.98 & m s$^{-2}$ & Acceleration from gravity \\
 R$_{\rm HJ}$ & 9.865$\cdot$10$^{7}$ & 9.865$\cdot$10$^{7}$ & 9.865$\cdot$10$^{7}$ & m & Radius of HJ \\
 $\Omega_{\rm HJ}$ & 2.063$\cdot$10$^{-5}$  & 2.063$\cdot$10$^{-5}$ & 2.063$\cdot$10$^{-5}$ & rad s$^{-1}$ & Rotation rate of HJ \\
 $\Delta$t$_{\rm hydo}$ & 30 & 30 & 20 & s & Hydrodynamic time-step \\
 $\Delta$t$_{\rm rad}$ & 30 & 30 & 100 &s & Radiaitve time-step \\
 N$_{\rm v}$ & 53 & 53 & 53 & - & Vertical level resolution \\
 d$_{\rm 2}$ & 0.02 & 0.02 & 0.02 & - & div. dampening coefficient \\
\hline
\end{tabular}
\label{tab:GCM_parameters}
\end{table*}

\section{Unbiased Monte Carlo RT transmission sampling}
\label{app:CMCRT}

In \citet{Lee2019} a transmission spectra mode in CMCRT was presented which was found to produce a biasing in the results due to binning of photon packets into discrete transmission limb impact parameters inside a spherical geometrical grid.
This led to a biasing towards higher optical depths, resulting in slightly higher R$_{p}$/R$_{\star}$ at the 10s ppm level compared to the benchmark cases.
Here we present a simple unbiased Monte Carlo sampling method for producing transmission spectra from GCM output.

The transmission spectra equation is given by \citep[e.g.][]{Dobbs-Dixon2013,Robinson2017}
\begin{equation}
\label{eq:trans}
\left(\frac{R_{p, \lambda}}{R_{\star}}\right)^{2} = \frac{1}{R_{\star}^{2}} \left(R_{p, 0}^{2} + 2\int_{R_{p, 0}}^{\infty}[1 - \mathcal{T}(b)]bdb\right),
\end{equation}
where R$_{p, \lambda}$ [m] is the wavelength dependent radius of the planet, R$_{\star}$ [m] the radius of the host star, R$_{p, 0}$ [m] the bulk planetary disk radius, $\mathcal{T}$ the transmission function, and $b$ [m] the impact parameter.
Formally the upper limit for the integral in Eq. \ref{eq:trans} is $\infty$.
This is replaced by the top of atmosphere radius, R$_{p, TOA}$ [m], as per the simulation output to facilitate numerical calculations.

Following the principles of integration through independent sampling, the result of the integral in Eq. \ref{eq:trans}, I$_{p}$, is approximated by simulating a suitably large number of N$_{ph}$ photon packets that sample the integral function
\begin{equation}
\langle I_{p}\rangle = \frac{(R_{p, TOA} - R_{p, 0})}{N_{ph}} \sum^{N_{ph}}_{i} [1 - e^{-\tau_{i}}]b_{i},
\end{equation}
where $\tau_{i}$ is the optical depth that the packet requires to escape the atmosphere toward the observational direction at impact parameter b$_{i}$.
The scheme therefore reduces to sampling a random impact parameter, b$_{i}$, between R$_{p, TOA}$ and R$_{p, 0}$ at randomly chosen transmission annulus for the photon packet.
The optical depths and impact parameter of each packet is then tracked and summed during the simulation, before normalisation at the end of the simulation.
In our testing, this scheme avoids the geometrical biasing found in \citet{Lee2019} at the cost of additional packet noise, requiring more packets ($\approx$10$\times$) to be simulated to reach a similar level of noise to the previous method.

We note this method is not limited to MCRT models but can be utilised by 1D/3D ray tracing codes to avoid manually choosing the initial start position of the optical depth tracing, this is the case for the \citet{Lee2019} which is a hybrid MCRT and ray tracing model.


\bsp	
\label{lastpage}
\end{document}